\documentclass[preprint2]{aastex631}

\usepackage{soul}
\usepackage{amsmath}
\setstcolor{red}

\shorttitle{$\gamma$ rays from LLAGN}
\shortauthors{Karwin et al.}

\graphicspath{{./}{Images/}}

\begin{document}

\title{Characterizing the $\gamma$-ray Emission from Low-Luminosity AGN}

\author[0000-0002-6774-3111]{Christopher M. Karwin}
\affiliation{Department of Physics and Astronomy, Clemson University, Clemson, SC 29634, USA}
\affiliation{NASA Postdoctoral Program Fellow, NASA Goddard Space Flight Center, Greenbelt, MD, 20771, USA}

\author[0009-0002-2068-3411]{Nikita S. Khatiya}
\affiliation{Department of Physics and Astronomy, Clemson University, Clemson, SC 29634, USA}

\author[0000-0003-1046-1647]{Margot Boughelilba}
\affiliation{Universität Innsbruck, Institut für Astro- und Teilchenphysik, 6020 Innsbruck, Austria}

\author[0000-0002-7791-3671]{Xiurui Zhao}
\affiliation{Cahill Center for Astrophysics, California Institute of Technology, Pasadena, CA 91125, USA}

\author[0000-0001-8604-7077]{Anita Reimer}
\affiliation{Universität Innsbruck, Institut für Astro- und Teilchenphysik, 6020 Innsbruck, Austria}

\author[0000-0002-6584-1703]{Marco Ajello}
\affiliation{Department of Physics and Astronomy, Clemson University, Clemson, SC 29634, USA}

\begin{abstract}

A majority of the active galactic nuclei (AGN) in the local Universe are classified as low-luminosity AGN (LLAGN), having bolometric luminosities $\lesssim 10^{42} \ \mathrm{erg \ s^{-1}}$. Although high-energy $\gamma$-ray emission is predicted from both the jets and
disks of LLAGN, to date only four have been detected by the \textit{Fermi} Large Area Telescope (\textit{Fermi}-LAT). In this work, we therefore
conduct a comprehensive study of all the LLAGN from the Palomar spectroscopic survey of bright, northern galaxies, including both subthreshold and detected $\gamma$-ray sources, using 14.4 years of LAT data. Our analysis results in a new detection of one LLAGN, as well as a detection of the subthreshold population using a stacking technique. We find that the signal from the subthreshold sample is consistent with being dominated by star-formation activity, although a contribution from compact jets or a mixed contribution from jetted and non-jetted systems is also feasible. On the other hand, the individually detected LLAGN are likely dominated by jet emission. We perform detailed spectral modeling for a subset of these sources and find that the $\gamma$-ray signal can be explained by synchrotron self-Compton radiation, if the inner jet emission region is weakly magnetized with its total energy density being strongly particle dominated, and only slowly moving. With this work we also publicly release our Python-based stacking library for analyzing subthreshold source populations with the LAT, based on a proven technique used in numerous studies.
\end{abstract}


\section{Introduction} \label{sec:intro}
Active galactic nuclei (AGN) are powered by the accretion of mass onto a supermassive black hole (SMBH). The accretion disk is conventionally considered to be geometrically thin and optically thick, with a mean accretion efficiency ($\eta$) of 10\%~\citep{2011ApJ...728...98D}; i.e., $\eta = L_{\mathrm{Bol}}/(\dot{M}_{\mathrm{BH}}c^2) = 0.1$, where $L_{\mathrm{Bol}}$ is the bolometric luminosity of the AGN, $\dot{M}_{\mathrm{BH}}$ is the mass accretion rate onto the SMBH, and $c$ is the speed of light. The infalling material loses a fraction of its rest mass energy, and the accretion efficiency gives the fraction of mass inflow converted into radiation. If all AGN in the local Universe ($d_L \leq 100$ Mpc) were in fact accreting with the canonical efficiency of 0.1, then there should be many nearby luminous sources, and this would still be the case even with very conservative estimates for the amount of available mass fueling the accretion~\citep{2009ApJ...699..626H}. However, we actually don't observe many nearby luminous sources. Rather, only $\sim1-5 \%$ of local galaxies contain bright Seyfert nuclei~\citep[e.g., see][and references therein]{2009ApJ...699..626H}. 

Most of the AGN in the local Universe are classified as low-luminosity AGN (LLAGN), having bolometric luminosities $\lesssim10^{42} \ \mathrm{erg \ s^{-1}}$~\citep{2009ApJ...699..626H,Saikia:2018tpp}. These sources are characterized by a low sub-Eddington accretion rate ($\dot{M}_{\mathrm{Edd}}$) of $\dot{M}_{\mathrm{BH}}<0.01\dot{M}_{\mathrm{Edd}}$~\citep{Ho:2008rf}. Consequently, a majority of LLAGN are thought to be in a radiatively inefficient accretion flow (RIAF) mode, where the radiative efficiency is much less than the canonical value of $\eta=0.1$~\citep{Ho:2008rf, 2009ApJ...699..626H}. In this mode the plasma cannot efficiently cool (its cooling time exceeds the accretion timescale), resulting in a high thermal pressure. This produces a geometrically thick, optically thin accretion disk. Particles in the plasma are accelerated to high energies, and nonthermal $\gamma$-ray emission may be generated. Additionally, observational evidence indicates that LLAGN are associated with relativistic jets, both compact and extended~\citep{2005A&A...435..521N,Baldi:2018uyo,Saikia:2018tpp}, which can also generate high-energy $\gamma$ rays. 

Physical quantities associated with accretion physics are thought to scale globally across BHs of all masses, from SMBHs in AGN to stellar-mass systems in X-ray binaries~\citep{Saikia:2015ega,Saikia:2018tpp,Baldi:2018uyo}. One indication of this is the empirical relation known as the fundamental plane of black hole activity (FPBHA), which suggests scale-invariance in accretion and jet production. Specifically, this relates the radio luminosity (a probe of the AGN jet), the X-ray or O\,{\sc iii} luminosity (a tracer of the accretion rate), and the BH mass, implying that the synchrotron radiation power emitted from a scale-invariant jet depends on the BH mass and the accretion rate, as given by the FPBHA. Both LLAGN and FR I radio galaxies have been observed to follow this relation.

Generally speaking, the classification of AGN is complex, as sources often meet multiple criteria across different schemes. In the radio domain, one way in which AGN are traditionally categorized is based on their radio power and the linear extent of their radio structure~\citep[see][and references therein]{2023A&ARv..31....3B}. These properties delineate distinct populations such as FR 0, FR I, and FR II radio galaxies, compact symmetric objects (CSOs), and compact steep-spectrum (CSS) sources, among others. LLAGN also occupy a distinct region within this parameter space, although their radio classifications can be diverse. For example, some resemble FR I radio galaxies, while others align more closely with CSOs. This heterogeneity in radio properties complicates efforts to generalize the $\gamma$-ray emission mechanisms of LLAGN, a topic further explored in this work.

From the $\gamma$-ray perspective, a strong correlation exists between the core radio luminosity (typically measured at 5 GHz) and the $\gamma$-ray luminosity for FR I and FR II radio galaxies~\citep{2011ApJ...733...66I,DiMauro:2013xta,2019ApJ...879...68S}. Recent studies show that this relation extends to even lower radio luminosities, encompassing the FR 0 population~\citep{2024ApJ...971...84K}. This suggests a common origin or environment for the emitting particle populations across these systems. It is therefore natural to ask whether a similar relationship holds for LLAGN more generally. Observations of $\gamma$-ray emission from LLAGN have the potential to provide valuable insight into the jet–disk connection and the physical processes driving jet formation in compact accreting systems.

To date, only one systematic study of the high-energy $\gamma$-ray emission from LLAGN has been conducted, using data from the \textit{Fermi} Large Area Telescope (\textit{Fermi}-LAT), which resulted in the significant detection of four sources~\citep{deMenezes:2020rah}. Two of these are very well known sources---NGC 4486 (commonly known as M87) and NGC 1275---and have already been well explained in terms of synchrotron and synchrotron self-Compton (SSC) jet models~\citep{2009ApJ...707...55A,2009ApJ...699...31A}. A detailed analysis of the spectral energy distributions (SEDs) was performed in~\citet{deMenezes:2020rah} for the two other sources (NGC 315 and NGC 4261), where a comparison was made between a RIAF model and a leptonic jet-dominated SSC model. Both SEDs were found to be better described by the SSC model. However, neither of the models was able to account for all of the observed $\gamma$-ray spectrum. Apart from the four reported detections, there remain 192 sources from the LLAGN sample\footnote{We use the same starting LLAGN sample that was used in~\citet{deMenezes:2020rah}, as described in Section~\ref{sec:sample}.} below the LAT detection threshold. Therefore, in this work we conduct a stacking study of the subthreshold sources, in order to characterize the average $\gamma$-ray properties of the LLAGN population. Additionally, we conduct SED modeling for a few of the significant sources, in order to better understand the emission mechanisms. Moreover, for the first time, we test for a scaling relation between the $\gamma$-ray luminosity and the infrared luminosity, as well as the core radio luminosity. 

Our analysis proceeds as follows. In Section~\ref{sec:sample} we discuss the sample selection. In Section~\ref{sec:data_analysis} we discuss the data analysis, which includes the data selection, background modeling, and stacking technique. Results are presented in Section~\ref{sec:results}. In Section~\ref{sec:model} we perform detailed SED modeling for three of the significant sources. Finally, in Section~\ref{sec:discussion_and_conclusion} we give our summary and conclusions. Throughout this analysis we assume the following cosmological parameters: $H_0 = 71 \ \mathrm{km \ s^{-1} \ Mpc^{-1}}$, $\Omega_\mathrm{m} = 0.27$, and $\Omega_\mathrm{\Lambda} = 0.73$.

\section{Sample Selection} 
\label{sec:sample}
Our source sample is selected from the Palomar survey\footnote{The full dataset from the Palomar survey is available at \url{https://vizier.cfa.harvard.edu/viz-bin/VizieR?-source=J/ApJS/112/315}} -- an optical spectroscopic survey of nuclei ($r \leq 200$ pc) of nearby galaxies conducted between 1984 and 1990~\citep{1995ApJS...98..477H,1997ApJS..112..315H,Ho:1997vg,1997ApJ...487..568H,ho2003search,2009ApJS..183....1H}. The survey consists of 486 galaxies from the Revised Shapley-Ames Catalog of Bright Galaxies (RSA) and the Second Reference Catalog of Bright Galaxies, satisfying the criteria $B_T\leq12$ mag and $\delta>0^\circ$, where $B_T$ is the total apparent magnitude in the B-band, and $\delta$ is the declination. The spectra were acquired using the Hale 5 m telescope at Palomar Observatory. 

The optical spectral range contains several emission lines whose intensity ratios can be used to discriminate different sources of ionization~\citep{1997ApJS..112..315H,Ho:2008rf}. This classification method was used to define four subclasses of the Palomar sample: H~II nuclei, low ionization nuclear emission line regions (LINERs), transition objects, and Seyfert nuclei. As discussed in the original studies, the H~II nuclei are assumed to be powered by the photoionization of young massive stars. The LINER emission appears to be non-stellar in origin. The optical emission line spectra broadly resemble those of traditional AGN such as Seyfert nuclei, but they have characteristically lower ionization levels. Finally, the transition objects are thought to be LINERs with contamination from nearby H II regions. The Seyferts, LINERs, and transition objects are assumed to be different classes of LLAGN.  

Our nominal subthreshold sample is comprised of all Seyferts, LINERs, and transition nuclei from the Palomar survey. We exclude the four sources that are already detected by the LAT (NGC 1275, NGC 4486, NGC 315, NGC 4261) and analyze them separately. In addition, our analysis results in a new detection of NGC 4374, and it is also excluded from the nominal subthreshold sample and analyzed separately. In order to avoid any possible confusion with emission from a blazar jet, we cross-correlate our LLAGN sample with the Roma-BZCAT\footnote{The Roma-BZCAT catalog is available at \url{https://heasarc.gsfc.nasa.gov/W3Browse/all/romabzcat.html}} catalog using an error radius of $0.1^\circ$. This results in four matches which are subsequently removed from the nominal subthreshold sample (NGC 676, NGC 3516, NGC 4698, and NGC 6503). In addition, we exclude NGC 4151, since the associated $\gamma$-ray emission is highly confused with emission from a known ultra-fast outflow, as well as a nearby blazar~\citep{2021ApJ...921..144A,Peretti:2023crf,2024ApJ...961L..34M}. This gives us a final sample of 186 subthreshold sources. 

\section{Data Analysis} \label{sec:data_analysis}
\subsection{Data Selection and Background Modeling} \label{sec:data_selection}
We analyze data collected by the LAT~\citep{Fermi-LAT09} from 2008 August 04 to 2023 January 05 (14.4 yr). For the subthreshold sample, we select gamma rays with energies in the range $1-800$ GeV, binned in 8 energy bins per decade. To reduce contamination from the Earth’s limb, we limit our selection to zenith angles no greater than 105$^\circ$. For each source, we deﬁne a 10$^\circ$ $\times$ 10$^\circ$ region of interest (ROI) centered at the optical position of the galaxy. The spatial bin size is 0.08$^\circ$. We select photons corresponding to the P8R3\_SOURCE\_V3 instrument response functions~\citep{Fermi-LAT13,Bruel+18}. In order to optimize the sensitivity of our analysis, we implement a joint likelihood fit with the four point-spread function (PSF) event types available in the Pass 8 data set\footnote{For more information on the different PSF types see~\url{https://fermi.gsfc.nasa.gov/ssc/data/analysis/documentation/Cicerone/Cicerone_Data/LAT_DP.html}.}. The data are divided into quartiles corresponding to the quality of the reconstructed direction, from the lowest quality quartile (PSF0) to the best quality quartile (PSF3). Since we use a binned likelihood analysis, each sub-selection has its own binned likelihood instance that is combined in a summed likelihood function for the ROI. 

Each PSF type has its own corresponding isotropic spectrum, namely, P8R3\_SOURCE\_V3\_PSF\emph{i}\_v1 (\emph{i} = 0$-$3). The Galactic diffuse emission is modeled using the standard component (gll\_iem\_v07\footnote{More information on the LAT background models can be found at~\url{https://fermi.gsfc.nasa.gov/ssc/data/access/lat/BackgroundModels.html}}), and the point-source emission is modeled using the 4FGL-DR3 catalog (gll\_psc\_v28)~\citep{2022ApJS..260...53A}, which is based on 12 years of data. To account for the additional exposure of the dataset compared to the source catalog, we also search for new point sources in each of the ROIs. In order to account for photon leakage from sources outside of the ROI due to the PSF of the detector, the model includes all 4FGL-DR3 sources within a 15$^\circ$ $\times$ 15$^\circ$ region. The energy dispersion correction (edisp\_bins = $-$1) is enabled for all sources except the isotropic component. We use the standard data ﬁlters: DATA$\_$QUAL $>$ 0 and LAT$\_$CONFIG==1.   

In addition to the nominal energy range, the subthreshold sample is also analyzed separately between $0.3 - 1$ GeV. In this case we use a maximum zenith angle of $100^\circ$, and all the other selections are the same as described above.

For the significant sources, we use similar selections as those used for the 4FGL-DR3 catalog \citep{2022ApJS..260...53A}. Specifically, we use different event types and zenith cuts depending on the energy interval, as summarized in Table~\ref{tab:data}. This was done in order to make a direct comparison with the catalog sources. The rest of the selections (i.e., number of energy bins per decade, spatial bin size, time range, and ROI size) are the same as described for the subthreshold sources. 
\begin{deluxetable}{lcc}
\tablecaption{Summed Likelihood Components}
\tablehead{
\colhead{Energy [GeV]} & \colhead{Z Max [deg]} & \colhead{PSF Types}    
}
\startdata
$0.05 - 0.1$    &80       &3 \\
$0.1 - 0.3$      &90       &2, 3 \\
$0.3 - 1$      &100       &1, 2, 3 \\
$1 - 1000$      &105       &0, 1, 2, 3 
\enddata
\label{tab:data}
\tablecomments{These selections are only used for the significant sources. The energy bins distinguish different components of the joint likelihood analysis; however, the underlying fits still use eight energy bins per decade, as described in the text. Z Max specifies the maximum zenith angle which controls the Earth limb contamination.}
\vspace*{-\baselineskip}
\vspace*{-\baselineskip}
\end{deluxetable}

\subsection{Stacking Method}
The subthreshold sources are analyzed using a stacking technique. This technique has been developed and applied previously for multiple studies, e.g.,~upper limits on dark matter interactions~\citep{lat_2011_dwarfs,2024PhRvD.109f3024M}, detection of the extragalactic background light~\citep{Ajello:2018sxm}, extreme blazars~\citep{paliya2019fermi}, star-forming galaxies~\citep{Ajello:2020zna}, ultra-fast outflows~\citep{2021ApJ...921..144A}, molecular outflows~\citep{2023ApJ...943..168M}, and FRO radio galaxies~\citep{2024ApJ...971...84K}. The analysis is performed using the fermi-stacking package\footnote{The fermi-stacking library is available at \url{https://fermi-stacking-analysis.readthedocs.io/en/latest/}}~\citep{karwin_2025_stacking}, a Python-based library which we publicly release with this work. The fermi-stacking library employs Fermipy (v1.2)\footnote{Fermipy is available at~\url{https://fermipy.readthedocs.io/en/latest/}}, which utilizes the underlying Fermitools (v2.2.0). 

The main assumption that we make for the stacking technique is that the sample can be characterized by average quantities like the average flux and the average photon index (when we model the galaxy spectra with a power law). Note that the stacking can also be performed for other parameters, such as luminosity. The underlying assumption relies on the population sharing a common emission mechanism that correlates with specific properties of the sources in the sample. In practice, the inferred parameters should be interpreted as an effective population average, reflecting intrinsic scatter in the properties across the sample. There are then two steps to the method. In the first step the model components are optimized for each ROI using a maximum likelihood fit. We evaluate the significance of each source in the ROI using the test statistic (TS), which is defined as:
 \begin{equation}\label{eq:TSeq}
     \mathrm{TS} = -2\mathrm{log}(\mathrm{L_0/L}),
\end{equation}
\noindent where $\mathrm{L_0}$ is the likelihood for the null hypothesis, and L is the likelihood for the alternative hypothesis. For the first iteration of the fit, the spectral parameters of the Galactic diffuse component (index and normalization) and the isotropic component are freed. In addition, we free the normalizations of all 4FGL sources with TS$\geq$25 that are within $5^\circ$ of the ROI center, as well as sources with TS$\geq$500 and within $7^\circ$. Lastly, the LLAGN source is fit with a power-law spectral model, and the spectral parameters (normalization and index) are also freed. In the first step we also find new point sources using the Fermipy function \textit{find\_sources}, which generates TS maps and identifies new sources based on peaks in the TS. The TS maps are generated using a power law spectral model with an index of $2.0$. The minimum separation between two point sources is set to $0.5^\circ$, and the minimum TS for including a source in the model is set to 16.

In the second step, 2D TS profiles are generated for the spectral parameters of each LLAGN source, where the TS is defined as in Eq.~\ref{eq:TSeq}. We scan photon indices from 1 to 4 with a spacing of 0.1 and total integrated photon flux (between 1--800 GeV) from $10^{-13}$ to $10^{-9}$ $\mathrm{ph \ cm^{-2}\ s^{-1}}$ with 40 logarithmically spaced bins, freeing just the parameters of the diffuse components. The choice of 40 bins provides sufficient sampling of the likelihood surface to accurately estimate the best-fit parameters and their uncertainties, while keeping the computational cost manageable. Note that the likelihood value for the null hypothesis is calculated at the end of the first step by removing the LLAGN source from the model. Since we perform a joint likelihood maximization in the different PSF event types (PSF0 $-$ PSF3), the total profile for each source is obtained by adding the profiles from each of the four event types. Lastly, the TS profiles for all sources are added to obtain the stacked profile.  The TS is an additive quantity, and so the stacked profile gives the statistical significance for the combined signal. For simplicity, we calculate the significance using a chi-squared distribution with two degrees of freedom. However, it should be noted that the stacked TS surface is obtained from a profile likelihood in which diffuse background parameters are re-optimized at each grid point. Additionally, the photon index is not defined under the null hypothesis of zero flux. Therefore, Wilks' theorem\footnote{Wilks' theorem states that, under standard regularity conditions, the likelihood ratio test statistic asymptotically follows a chi-square distribution with degrees of freedom equal to the difference in the number of free parameters between the competing models.} does not strictly apply to the full two-parameter scan, and the quoted significance should be interpreted as approximate. 

We validated our stacking analysis previously with realistic Monte Carlo simulations of the \textit{Fermi}-LAT sky that include diffuse backgrounds and an unresolved blazar population, as described in~\citet{2021ApJ...921..144A}. These tests show that the method does not produce spurious detections when applied to empty sky positions, and that it accurately recovers the average flux and spectral index of a simulated source population, demonstrating that the stacking procedure robustly measures the mean properties of faint, below-threshold sources. We further validate our results in this work by repeating the analysis in blank sky regions, as detailed in Section~\ref{sec:controls}.

\section{Results}
\label{sec:results}
\subsection{Flux-Index Stack for Subthreshold Sources}

\begin{figure}[t]
\begin{center}
Full Sample 
\includegraphics[width=0.45\textwidth]{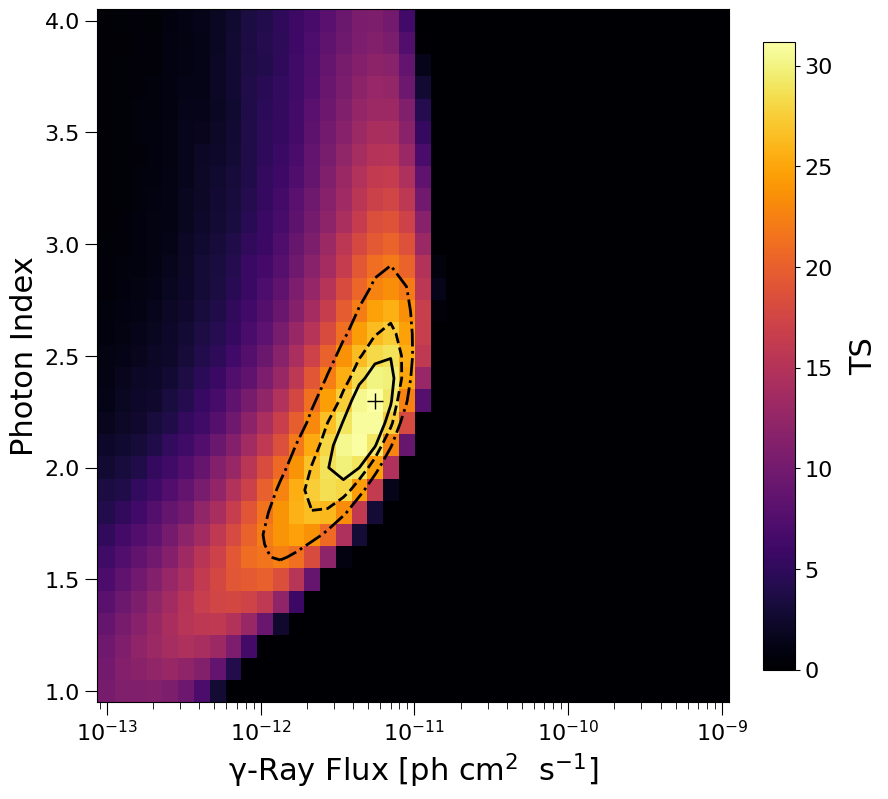}

\caption{Stacked TS profile for the sample of subthreshold LLAGN. The color scale indicates the TS, and the plus sign indicates the location of the maximum value, with a $\mathrm{TS} = 31.2$ (5.2\,$\sigma$). Significance contours (for 2 degrees of freedom) are overlaid on the plot showing the 68\%, 90\%, and 99\% confidence levels, corresponding to $\Delta\mathrm{TS}$ = 2.30, 4.61, and 9.21, respectively.}
\vspace*{-\baselineskip}
\label{fig:flux_index_stack}
\end{center}
\end{figure}

The results from the flux-index stacking are shown in Figure~\ref{fig:flux_index_stack}. The sample is significantly detected, with a maximum TS of 31.2, corresponding to a significance of roughly 5.2\,$\sigma$ (for 2 degrees of freedom). The best-fit flux (integrated from $1-800$ GeV) is $5.5^{+3.3}_{-2.1} \times 10^{-12} \ \mathrm{ph \ cm^{-2} \ s^{-1}}$, and the best-fit index is $2.3^{+0.2}_{-0.3}$. Note that the errors reported here do not account for the covariance between the variables; instead, they are the 1\,$\sigma$ errors with respect to the best-fit values. An indication of the parameter correlation can be inferred from the skewness of the significance contours, as shown in Figure~\ref{fig:flux_index_stack}. 

\begin{figure}
\begin{center}
\includegraphics[width=0.45\textwidth]{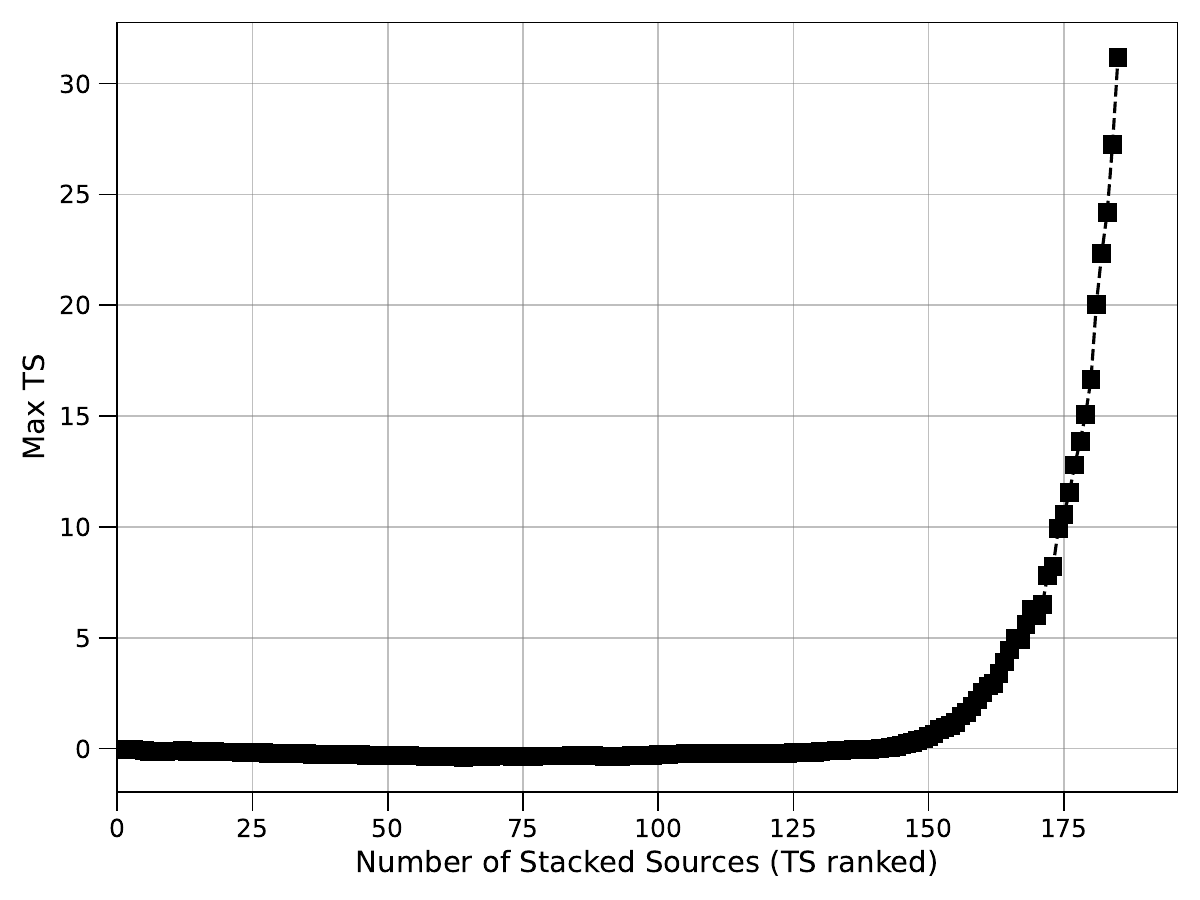} 
\caption{Maximum TS as a function of number of stacked sources, where the sources are ranked in order of increasing TS.}
\vspace*{-\baselineskip}
\label{fig:evolution_plot}
\end{center}
\end{figure}

Whenever performing a stacking analysis, it is important to verify that the signal is not just dominated by a few bright sources. In order to do this, we plot the maximum TS as a function of the number of stacked sources, where the sources are ranked in order of increasing TS. This is shown in
Figure~\ref{fig:evolution_plot}. As can be seen, $\sim40$ sources contribute positively to the overall TS. We can therefore conclude that the signal is not being dominated by just a few sources.

\begin{figure*}
\begin{center}
\includegraphics[width=0.48\textwidth]{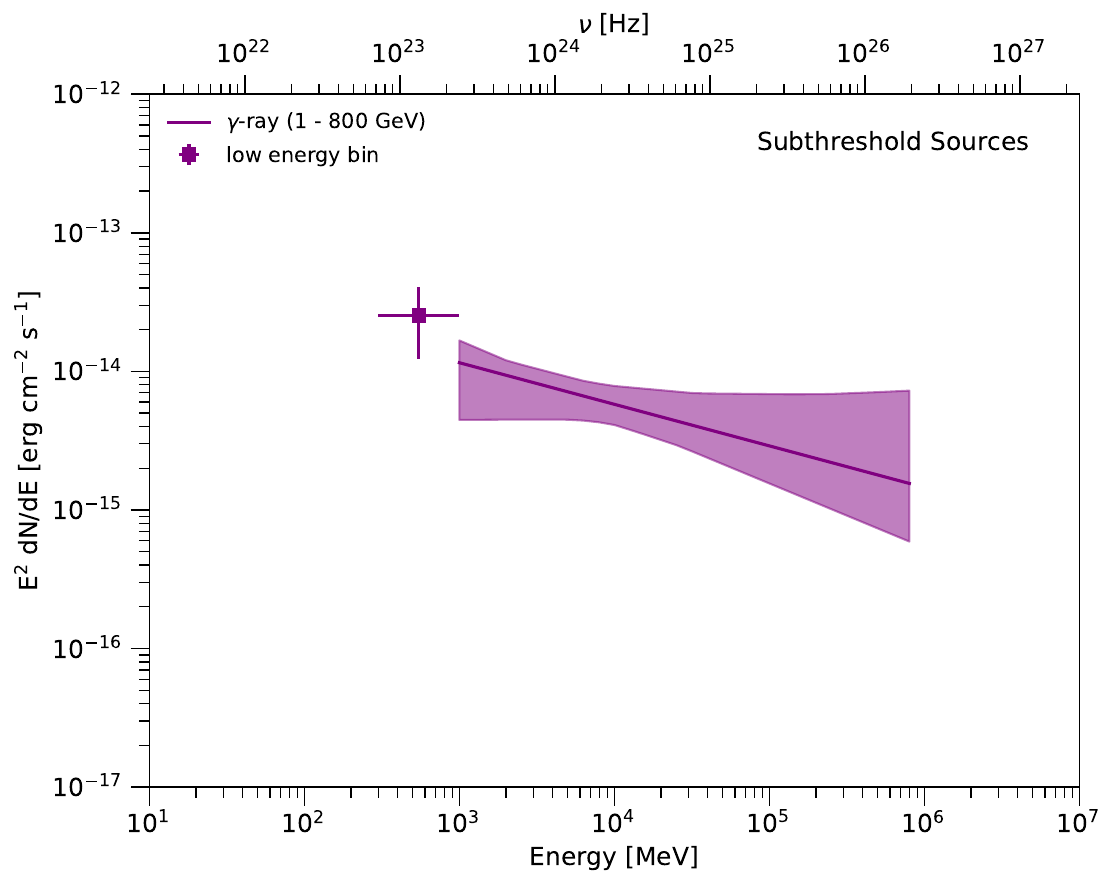} 
\includegraphics[width=0.48\textwidth]{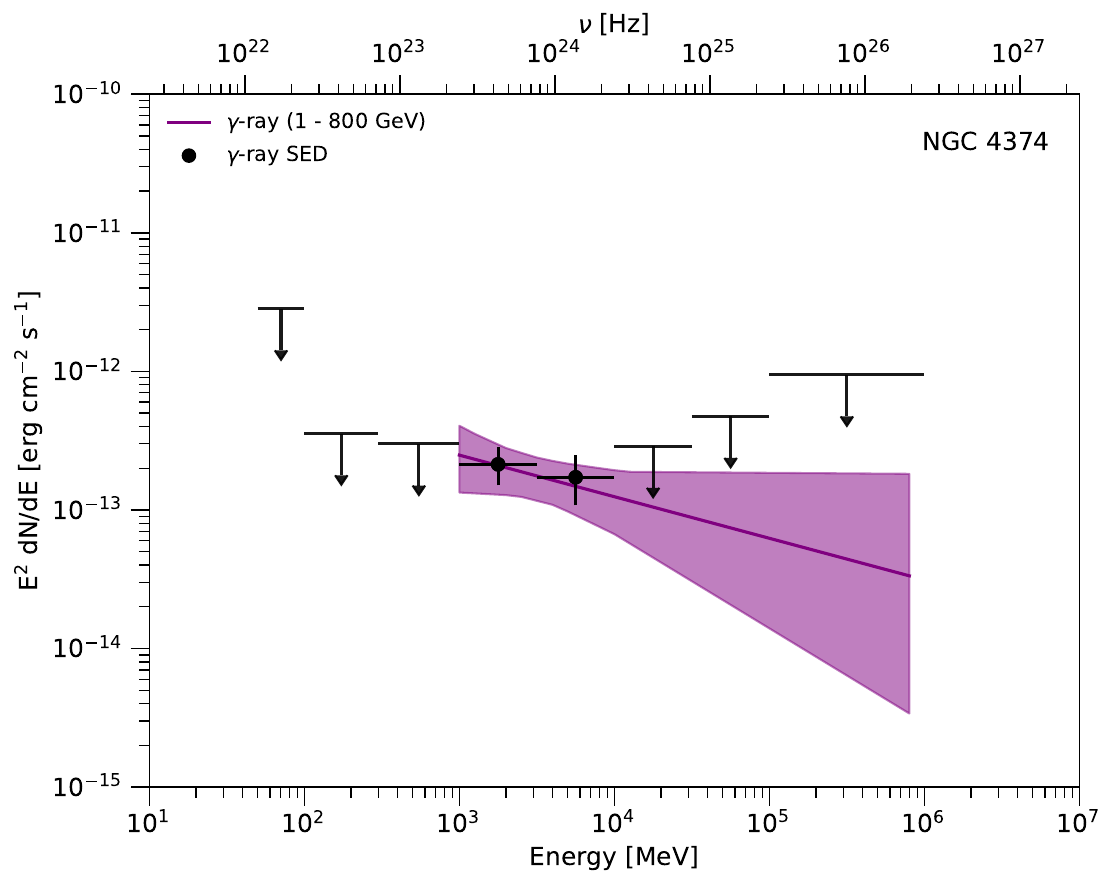} 
\includegraphics[width=0.48\textwidth]{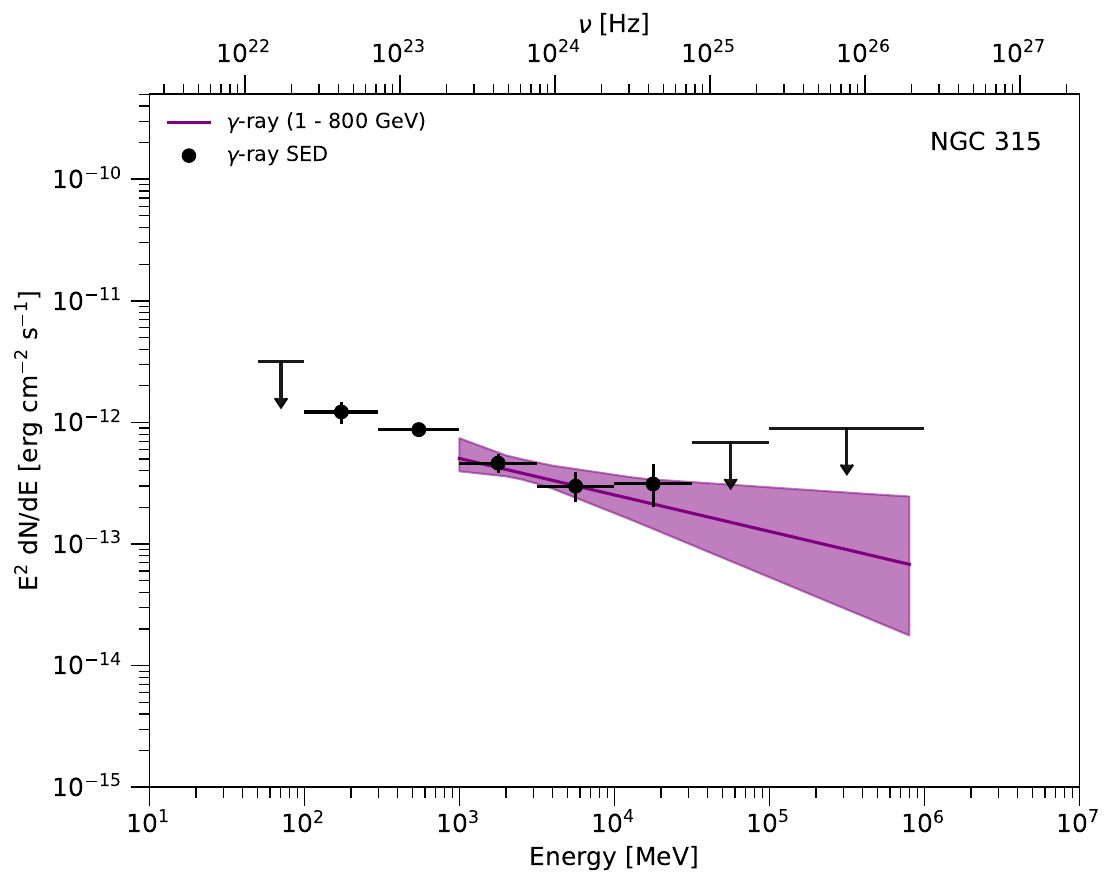} 
\includegraphics[width=0.48\textwidth]{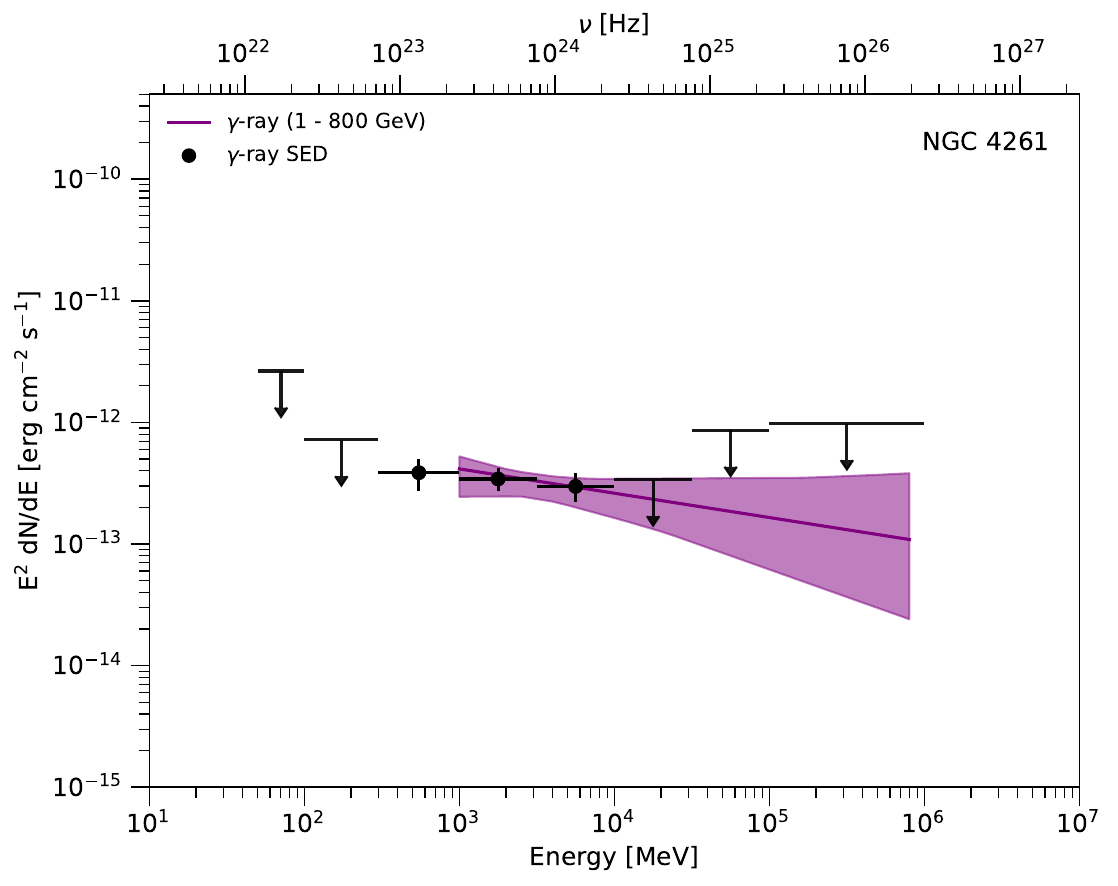} 
\includegraphics[width=0.48\textwidth]{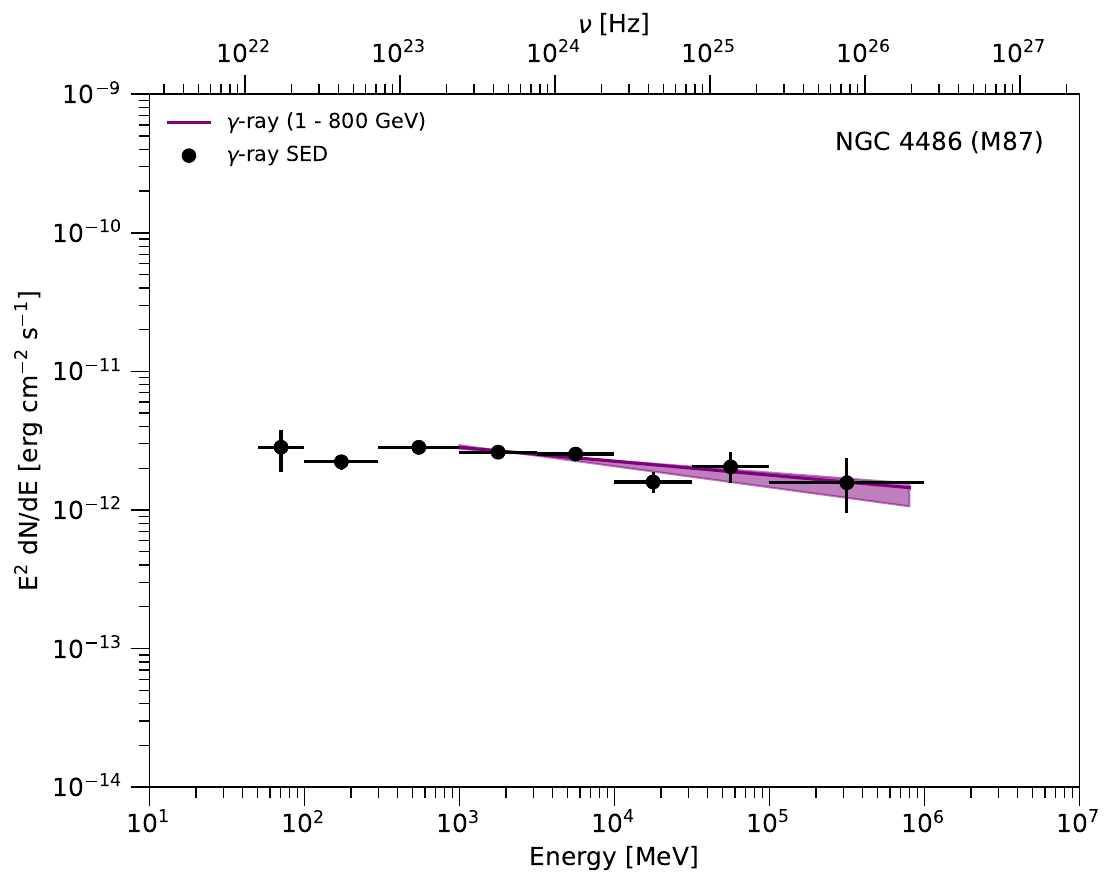} 
\includegraphics[width=0.48\textwidth]{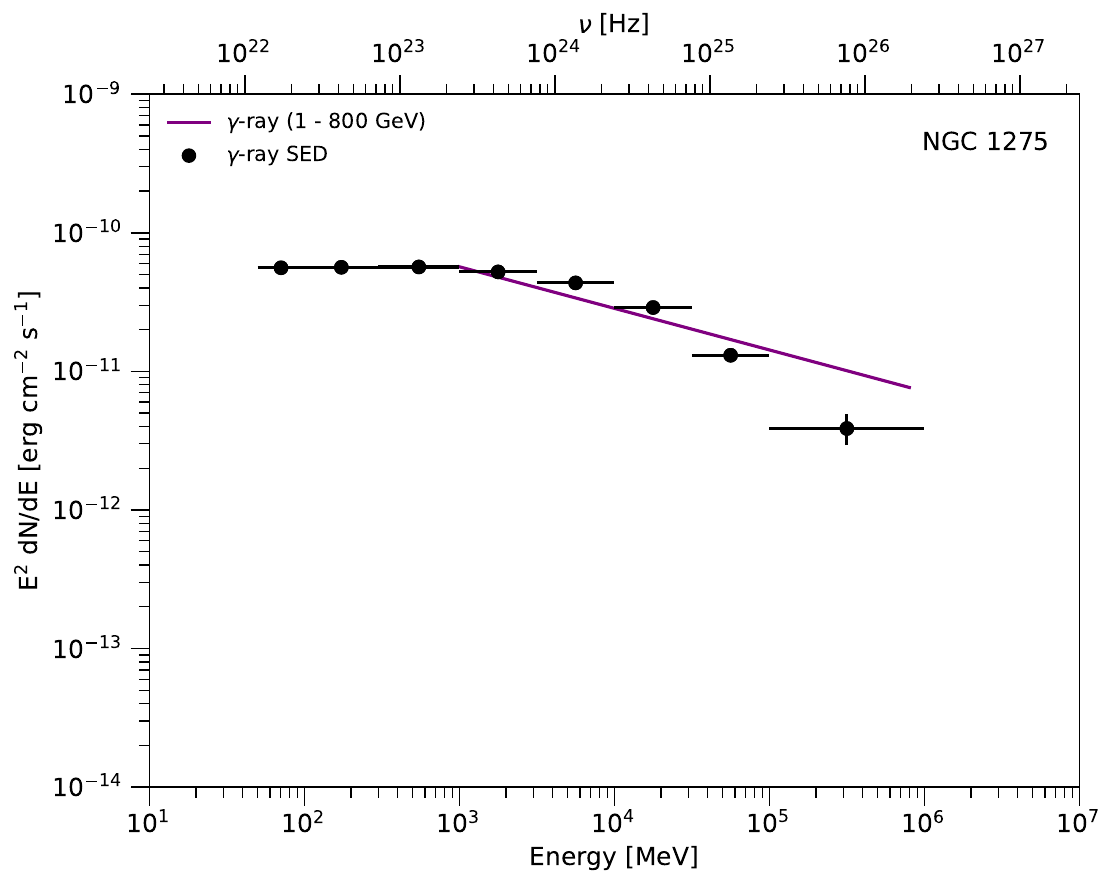} 
\caption{SEDs (black data points) and butterfly plots (purple bands) for the subthreshold sources and the significant sources, as specified in the legends. The error for the flux data points and the bands is at the 1\,$\sigma$ confidence level. Upper limits for the SEDs are plotted for bins with TS$<$9, and they are shown at the 95\% confidence level. The lower x-axis gives the energy, and the upper x-axis gives the frequency.}
\label{fig:seds}
\end{center}
\end{figure*}

We obtain the spectrum of the subthreshold sources by sampling spectral parameters within the 1\,$\sigma$ confidence region of the stacked profile. The corresponding butterfly plot is shown in the upper left panel of Figure~\ref{fig:seds}. In order to characterize the behavior at low energy, we repeat the stacking analysis for energies between 300 MeV $-$ 1 GeV. We use the same selections as for the nominal energy range, with the exception that we use a maximum zenith angle of $100^\circ$. We find a fairly bright signal in the low-energy bin, with a maximum TS of 14.5 (3.4\,$\sigma$) at a flux (integrated from $0.3-1$ GeV) of $3.7^{+2.2}_{-1.9} \times 10^{-11} \ \mathrm{ph \ cm^{-2} \ s^{-1}}$. The spectral index is not well constrained, so we use a value of 2.0, which is standard for SED calculations. 

As described in Section~\ref{sec:sample}, we cross-correlate our sample with the Roma-BZCAT catalog to mitigate contamination from blazar jets, though this catalog is not exhaustive. As a further check, we additionally cross-correlate with the CRATES catalog~\citep{2007ApJS..171...61H} of flat-spectrum radio sources and the Yuan and Wang catalog~\citep{2012ApJ...744...84Y} of radio-loud AGN, identifying 10 additional associations. Excluding these sources does not significantly affect our results, reinforcing that the signal is not driven by blazar contamination.

\subsection{Significant Sources}
As an initial step in our stacking pipeline, we check for any new sources that are significantly detected (i.e.,~TS$\geq$25). We find one new source, NGC 4374, which is detected for the first time, having a TS of 31.4. This is only the fifth LLAGN significantly detected in $\gamma$ rays. We calculate SEDs for all significant sources between 50 MeV $-$ 1 TeV, using the selections described in Section~\ref{sec:data_selection}. These are shown in Figure~\ref{fig:seds}. For the sources that were previously detected, we made a direct comparison with the SEDs reported in the 4FGL-DR3 and verified that our calculations are in excellent agreement for all sources and all energy bins. We calculate upper limits for bins with TS$<$9, using a frequentist approach (as implemented in Fermipy). However, this method is generally not applicable when the TS$\lesssim$1, in which case we use a Bayesian approach. This is a similar strategy as that employed for the 4FGL catalog~\citep{4FGL}. In addition to the SEDs, we calculate butterfly plots between 1 $-$ 800 GeV, which are also shown in Figure~\ref{fig:seds}. As can be seen, there is excellent consistency between the SEDs and the butterfly plots, as expected.

\begin{figure}
\begin{center}
Seyferts \\ 
\includegraphics[width=0.40\textwidth]{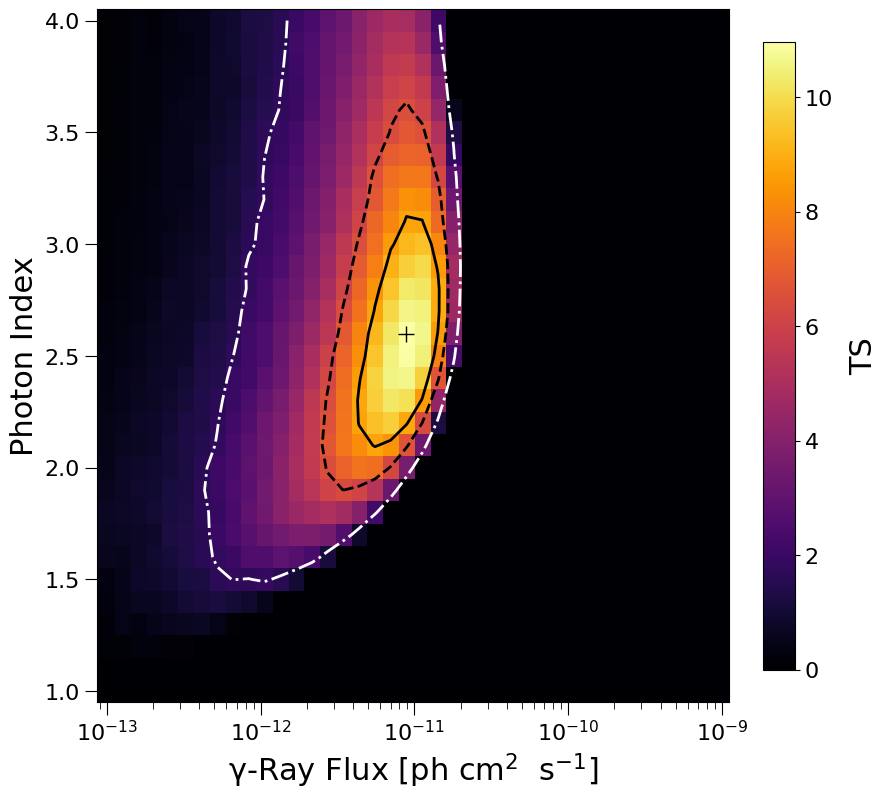} 
LINERs \\ 
\includegraphics[width=0.40\textwidth]{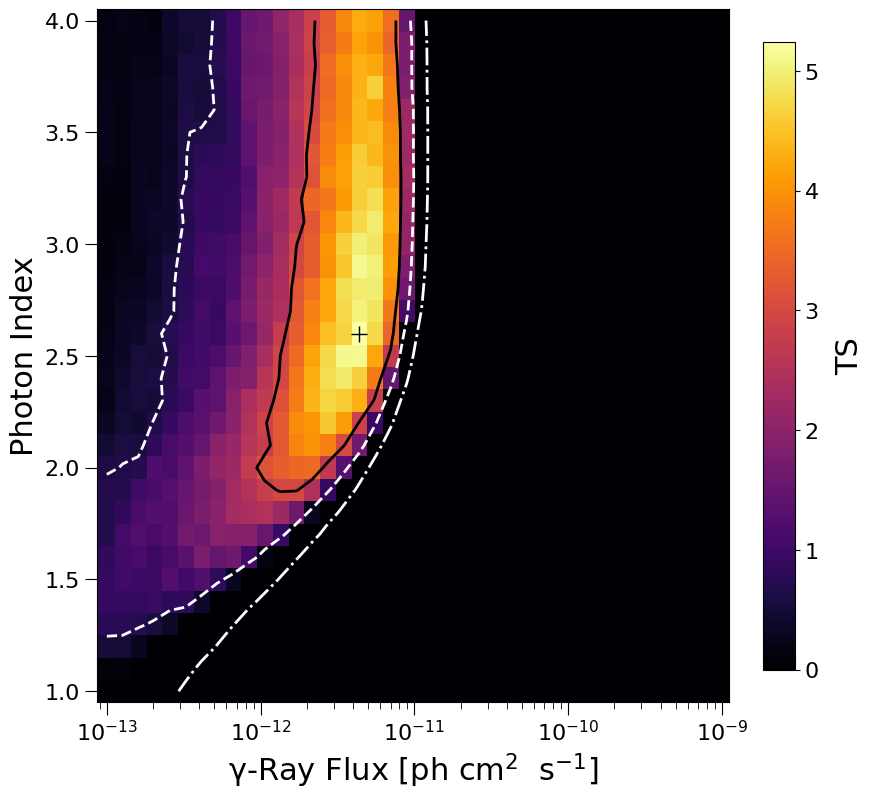}
Transition Nuclei \\
\includegraphics[width=0.40\textwidth]{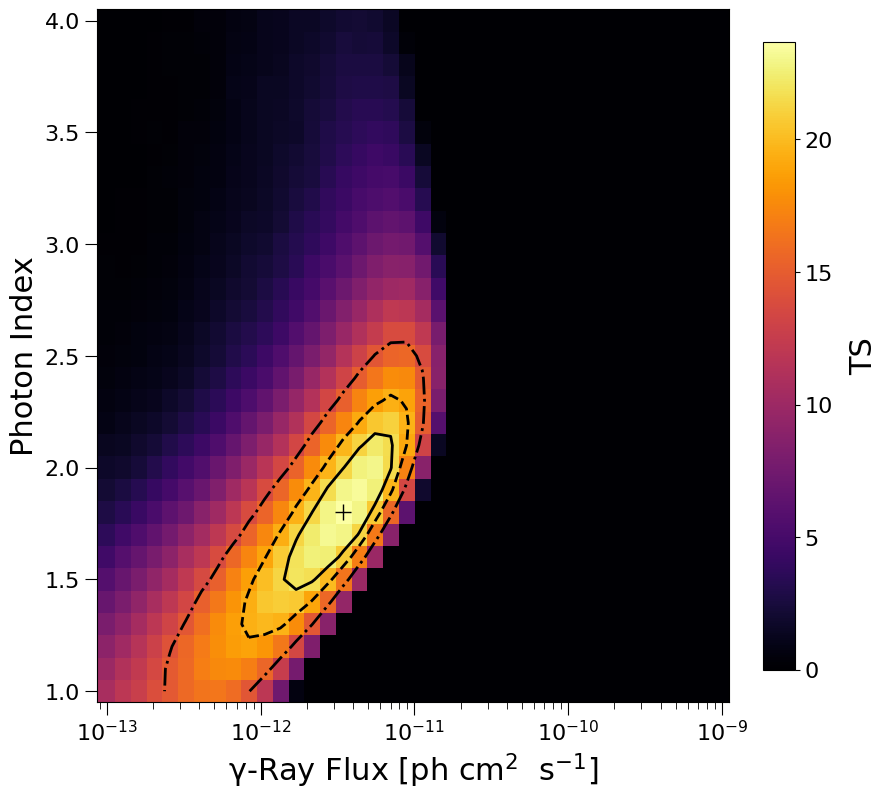} 
\caption{Stacked profiles for different subsets of the full sample: Seyferts, LINERs, and transition nuclei. The color scale and contours are the same as described in Figure~\ref{fig:flux_index_stack}. For visibility, contours are shown in white when needed.}
\label{fig:components}
\end{center}
\end{figure}

\subsection{Physically Motivated Subsets}


\begin{figure*}
\begin{center}
\begin{tabular}{c c}
Spirals & Non-Spirals  \\
\includegraphics[width=0.4\textwidth]{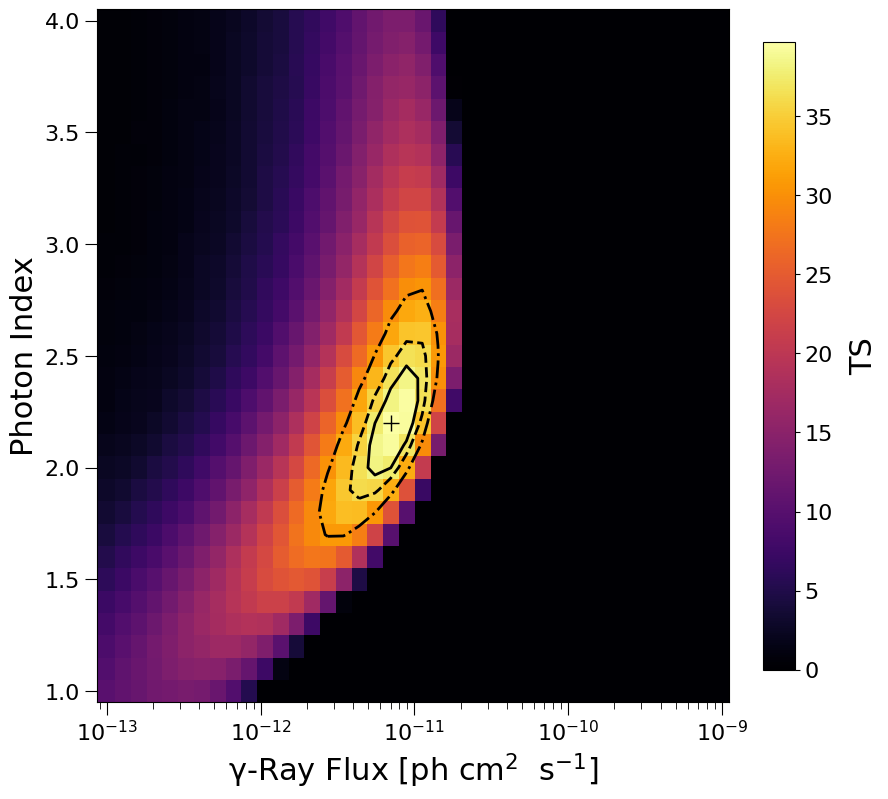} 
& \includegraphics[width=0.4\textwidth]{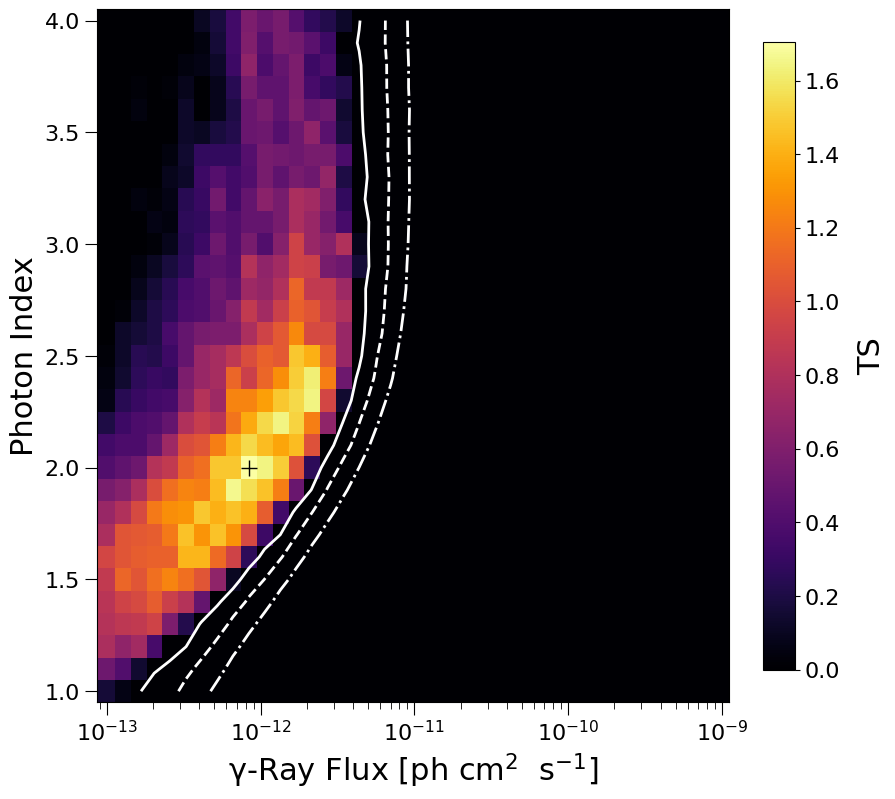} 
\end{tabular}
\caption{Stacked profiles for different subsets of the full sample: spirals and non-spirals. The color scale and contours are the same as described in Figure~\ref{fig:flux_index_stack}. For visibility, contours are shown in white when needed.}
\label{fig:spirals}
\end{center}
\end{figure*}

We can gain some physical insight into the nature of the signal by
considering different physically motivated subsets of the full sample. In
Figure~\ref{fig:components}, we show separate TS profiles for the
Seyferts, LINERs, and transition nuclei. The corresponding best-fit
spectral parameters are reported in
Table~\ref{tab:flux_index_results}. These three subsets are detected
at the significance levels of 2.9\,$\sigma$, 1.8\,$\sigma$,
and 4.5\,$\sigma$, respectively. 

LINERs are generally thought to be ``dwarf Seyferts", i.e.,~the nuclear emission is powered by accretion onto a SMBH. This interpretation is particularly clear for LINERs for which broad emission lines have been detected, as this indicates an underlying physical commonality with other AGN. This is the case for 18/84 LINERs in our sample. However, the nature of LINERs without detected broad emission lines is less clear. From the $\gamma$-ray observations, we find that the spectral parameters of the Seyferts and LINERs are consistent within 1\,$\sigma$ uncertainty. Regarding the maximum likelihood values, the index is the same for both subsets, whereas the flux from the Seyferts is $\sim2$x higher. These results seem to be in line with the general interpretation of LINERs being ``dwarf Seyferts". However, we are unable to make any robust conclusions from the $\gamma$-ray observations alone, as the overall significance is too low. That said, if we further divide the LINERs into sources with and without broad (H$\alpha$) emission lines, we find that the former are detected at a significance level of 2.3\,$\sigma$, whereas the latter are essentially undetected (0.8\,$\sigma$). While this finding is worth noting, we again cannot draw any definitive conclusions due to the overall low significance of the signal. 

\begin{deluxetable*}{lccccc}
\tablecaption{Flux-Index Stacking Results} \label{tab:flux_index_results}
\tablehead{
\colhead{Sample} & \colhead{$N$} & \colhead{TS} & \colhead{$\sigma$} & \colhead{Flux ($\times10^{-12}$)}  & \colhead{$\Gamma$}  \\
\colhead{}  & \colhead{} & \colhead{} & \colhead{} & \colhead{[$\mathrm{ph \ cm^{-2} \ s^{-1}}$]} & \colhead{} 
}
\startdata
Full        & 186 &31.2 & 5.2 &$5.5^{+3.3}_{-2.1}$ & $2.3^{+0.2}_{-0.3}$ \\
Transitions & 62  &23.7 & 4.5 &$3.5^{+2.1}_{-1.3}$ & $1.8^{+0.2}_{-0.2}$ \\
Seyferts    & 40  &11.0 & 2.9 &$8.9^{+5.4}_{-4.5}$  & $2.6^{+0.6}_{-0.5}$ \\
LINERs      & 84  &5.2  & 1.8 &$4.4^{+4.5}_{-3.0}$ & $2.6^{+0}_{-0.4}$ \\
LINERs (w/ H$\alpha$)     & 18  &7.5  & 2.3 &$7.0^{+7.2}_{-4.9}$ & $2.2^{+0.7}_{-0.4}$ \\
LINERs (w/o H$\alpha$)      & 66  &1.7  & 0.8 &\nodata & \nodata \\
Spirals     & 107 &39.7 & 6.0 &$7.0^{+4.2}_{-1.5}$  & $2.2^{+0.2}_{-0.3}$ \\
Non-Spirals & 79  &1.7  & 0.8 &\nodata  & \nodata 
\enddata
\tablecomments{The full sample is comprised of Transition Nuclei, Seyferts, and LINERs. $N$ is the number of sources in the stacked profile, $\sigma$ is the statistical significance, calculated for two degrees of freedom, and $\Gamma$ is the spectral index. Note that the errors reported here do not account for the covariance between the variables; rather, they are the 1\,$\sigma$ errors with respect to the best-fit values.}
\vspace*{-\baselineskip}
\end{deluxetable*}
Transition objects are generally thought to be LINERs with contamination from nearby H~II regions. We find that the maximum likelihood index of the transition objects is much harder than that of the LINERs and Seyferts ({see Table~\ref{tab:flux_index_results}), and there is also a stronger positive correlation between the parameters. However, within 1\,$\sigma$ uncertainty, the spectral parameters of all three subsets are consistent (particularly when we take into account the parameter covariance). The reason for the preference of a harder spectral index is not entirely clear. It could plausibly be due to contamination from nearby H~II regions, although when performing the stacking for H~II sources in the Palomar survey we find a maximum likelihood spectral index close to 3.0.

As another test, we divide the sample into categories based on the morphology of the galaxies, comparing spirals to non-spirals (i.e.,~ellipticals and lenticulars). This classification is based on the numerical Hubble stage, which is a parameter that is provided for all sources in the Palomar survey. The stacked profiles for the spiral and non-spiral cases are shown in Figure~\ref{fig:spirals}, and the corresponding best-fit spectral parameters are reported in Table~\ref{tab:flux_index_results}. The spiral galaxies are detected at a significance of 6.0\,$\sigma$. This is greater than for the full sample, even though there are only 107 sources in the subset. On the other hand, the non-spirals have essentially no signal. Thus, we can conclude that all the signal is essentially coming from galaxies with a spiral morphology. Intuitively, the non-spirals are more early-type galaxies compared to the spirals. The AGN may be less active in these early-type systems, and thus there is a weaker $\gamma$-ray signal. At the same time, the spirals likely have more star-formation activity (SFA) compared to the non-spirals, which must be taken into careful consideration when interpreting the $\gamma$-ray signal. 

\subsection{$\gamma$ Rays from Star-Formation Activity}
SFA is a well-known source of $\gamma$-ray emission. Indeed, for galaxies there exists a robust scaling relationship between $\gamma$-ray luminosity and SF rate, as traced by the total infrared (IR) luminosity between $8 - 1000 \ \mu \mathrm{m}$~\citep{2020ApJ...894...88A}. In this section we test if SF can plausibly account for the observed $\gamma$-ray signal from the sample of subthreshold LLAGN. Specifically, we test the relation
\begin{equation}
\label{correlation}
    \mathrm{log \ {L_\gamma}}=\beta + \alpha \ \mathrm{log}\left(\frac{L_{8-1000 \ \mu \mathrm{m}}}{10^{43.6} \ \mathrm{erg \ s^{-1}}}\right), 
\end{equation} 
where  $\alpha$ and $\beta$ are the scanned parameters in the fit. The quantity $L_{8-1000 \ \mu \mathrm{m}}$ is the IR luminosity in the wavelength range from 8-1000 $\mu \mathrm{m}$ calculated from its corresponding flux ($F_{8-1000 \ \mu \mathrm{m}}$), which is determined by a weighted sum of flux densities collected from the IRAS point source catalog~\citep{1988iras....7.....H}:

\begin{equation} \label{F_IR}
\begin{split}
      F_{8-1000 \ \mu \mathrm{m}} = 1.8 \times 10^{-14} (13.48 f_{12} + 5.16 f_{25} + \\ 2.58 f_{60} + f_{100}),
\end{split}
\end{equation}
where $f_{12}, \ f_{25}, \ f_{60}, \ \mathrm{and} \ f_{100}$ are the flux densities as 12, 25, 60, and 100 $\mu \mathrm{m}$, respectively. The normalization value of $10^{43.6} \mathrm{\ erg \ s^{-1}}$ is roughly equal to the mean IR luminosity of the sample. The $\gamma$-ray luminosity, $\mathrm{L_{\gamma}}$, is calculated between 1 - 800 GeV using the expression: 
\begin{equation}
\label{flux-lum}
    L_\gamma = 4  \pi  d_L^2 (z) \ \frac{F_{\gamma}}{(1 + z)^{2 - \Gamma}}, 
\end{equation}
where $d_L^2 (z)$ is the luminosity distance at redshift z, $(1 + z)^{2 - \Gamma}$ is the K$-$correction factor (with $\gamma$-ray index, $\Gamma$), and $F_{\gamma}$ is the energy flux integrated over the range $1 - 800$ GeV. 


\begin{figure}
\begin{center}
Infrared Correlation 
\includegraphics[width=0.45\textwidth]{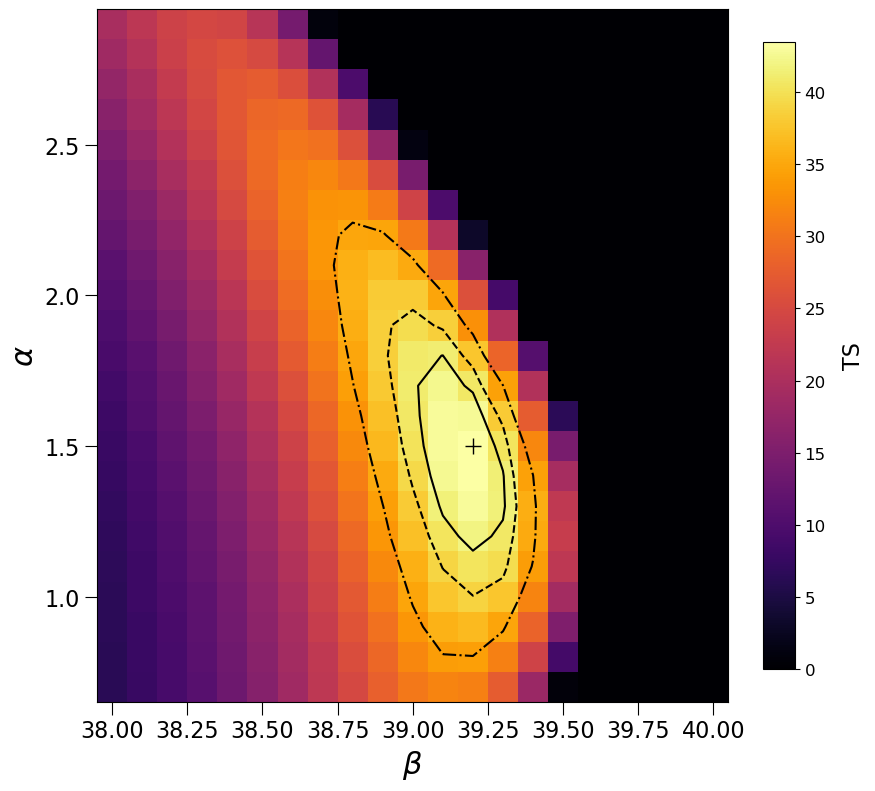} 

\caption{Profile from $\mathrm{L_{\gamma} - L_{8-1000 \mu m}}$ stacking for subset of spiral galaxies with IR data (98 sources). The color scale and contours are the same as described in Figure~\ref{fig:flux_index_stack}.}
\label{fig:SF_all_spirals}
\end{center}
\end{figure}
We scan $\alpha$ values between $0.7 - 2.9$ and $\beta$ values between $38 - 40$, both having a step size of 0.05. For converting the flux, we use the best-fit spectral index from the flux stacking. The stack includes 122 sources, based on the availability of IR data from the IRAS catalog. We get best-fit parameters of $\alpha$ = 1.4$_{-0.35}^{+0.35}$ and $\beta$ = 39.2$_{-0.15}^{+0.15}$, with a maximum TS of 44.9, corresponding to a significance of 6.4\,$\sigma$, for two degrees of freedom. If we restrict ourselves to just the spiral galaxies from this subset (98 sources), we get similar results, as shown in Figure~\ref{fig:SF_all_spirals}, with best-fit parameters of $\alpha$ = 1.5$_{-0.4}^{+0.2}$ and $\beta$ = 39.2$_{-0.2}^{+0.1}$, and a maximum TS of 43.5, corresponding to a significance of 6.3\,$\sigma$. This again shows that the signal is mostly coming from the spiral galaxies.

We compare our results to the established scaling relation for star-forming galaxies (SFGs) from~\citet{ajello2020}, shown in Figure~\ref{LLAGN_SFG_LgamvsLIR}. The original relation is determined using a $\gamma$-ray energy range of 100 MeV $-$ 800 GeV, and so we scale the relation to match the energy range used in our analysis. This comparison shows that the $\gamma$-ray luminosity of subthreshold LLAGN scales with the IR luminosity in a similar fashion as the SFGs, agreeing within the 1\,$\sigma$ uncertainties. Thus, the $\gamma$-ray emission from the subthreshold LLAGNs is consistent with being dominated by SFA. However, it is feasible that there may be a contribution from compact jets in these sources, as we explore further in the next section.

\begin{figure}[t]
\begin{center}
 
\includegraphics[width=0.49\textwidth]{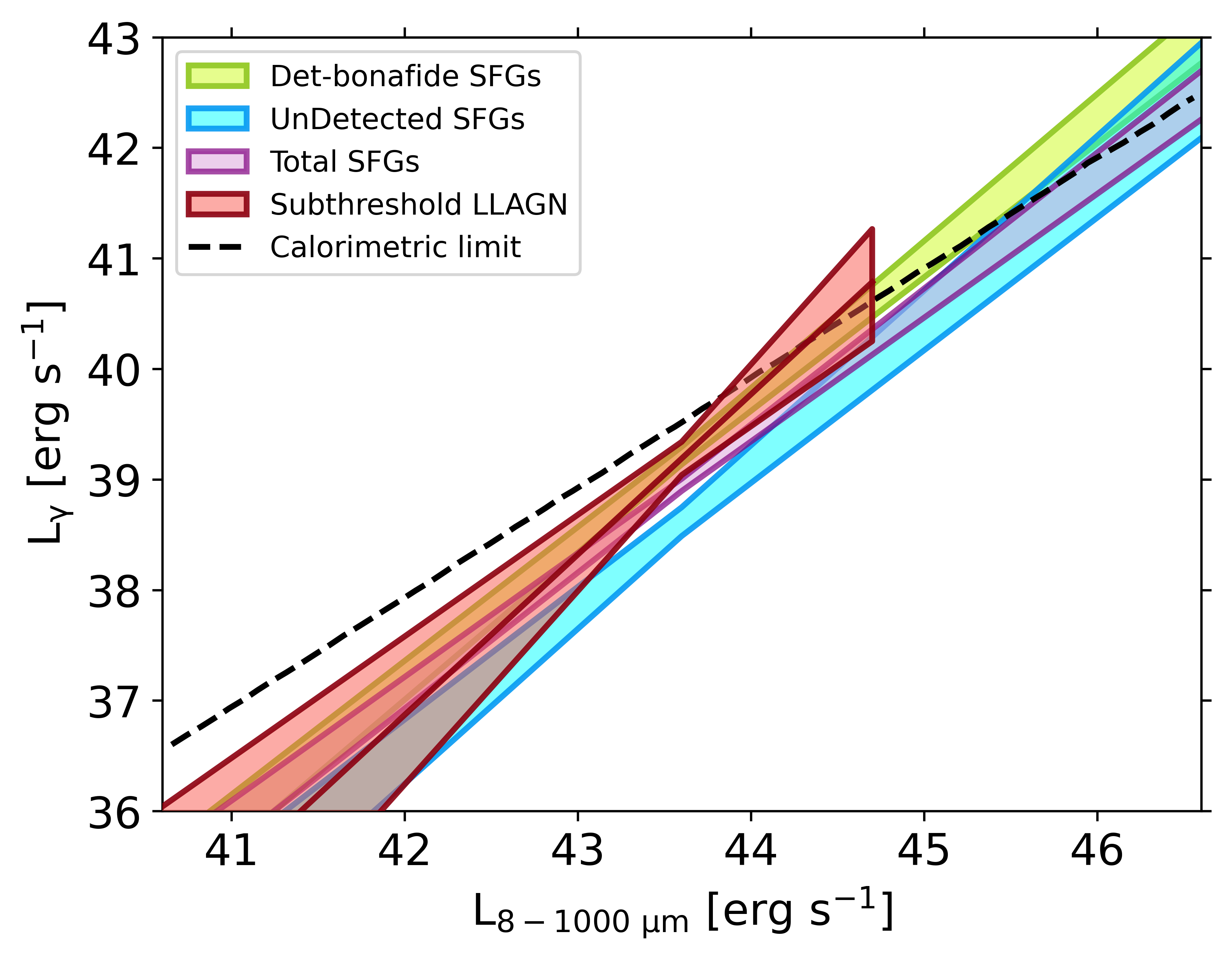}
 \caption{The $\mathrm{L_{\gamma} - L_{8-1000 \mu m}}$ correlation for subthreshold LLAGN is shown with the red band. The bonafide detected, undetected, and total SFGs from~\citet{ajello2020} are indicated in green, blue, and purple, respectively. All bands are at the 1\,$\sigma$ confidence level. The black dashed line shows the calorimetric limit. Both axes are in log scale.}
\label{LLAGN_SFG_LgamvsLIR}
\end{center}
\end{figure}

\subsection{$\gamma$ Rays from Compact Jets}
\label{subsec:Luminosity Correlation Study}

LLAGN have been associated with relativistic jets, both compact and extended~\citep{2005A&A...435..521N,Baldi:2018uyo,Saikia:2018tpp}, which can generate high-energy $\gamma$ rays. Indeed, there is a well-established correlation of the $\gamma$-ray luminosity ($L_\gamma$) with the core radio luminosity ($L_R$) for radio galaxies (FR~0, FR~I, and FR~II)~\citep[e.g.,][]{2011ApJ...733...66I, DiMauro:2013xta, 2019ApJ...879...68S,2024ApJ...971...84K}. We therefore test for a similar relation with the subthreshold LLAGN, which takes the form 
\begin{equation} \label{Eq:alpha-beta}
    \mathrm{log}L_\gamma =  \beta +   \alpha\mathrm{log}\bigg(\frac{L_{15\mathrm{GHz}}}{10^{38.2} \mathrm{\ erg \ s^{-1}}}\bigg),
\end{equation}
where the value in the denominator is the mean radio luminosity of the sample, $\alpha$ gives the slope of the relation, and $\beta$ is the $\gamma$-ray luminosity at the value of the mean radio luminosity. For the radio observations, we use the 15 GHz radio data from~\citet{2005A&A...435..521N} and~\citet{Saikia:2018tpp}. For the LLAGN sample, 129 of the sources were detected in radio, whereas for the remaining sources, we only have upper limits. We therefore perform the stacking using the subset of 129 sources detected at 15 GHz.

The results of stacking in $\mathrm{\alpha - \beta}$ space are $\alpha = 0.55_{-0.1}^{+0.1}$ and $\beta = 39.6_{-0.15}^{+0.2}$, with a maximum TS of 37.5, corresponding to 5.8\,$\sigma$, as shown in Figure \ref{fig:AB_Lgam_L15GHz_all_spirals}. We also study this correlation for the spirals-only case of 77 sources. The results from the stacked $\alpha - \beta$ profile show $\alpha = 0.5_{-0.15}^{+0.15}$ and $\beta = 39.65_{-0.19}^{+0.19}$, with a maximum TS of 24.3, corresponding to 4.5\,$\sigma$. We thus see that a significant correlation is found between $L_{\gamma}$ and $L_{15\mathrm{GHz}}$, and this correlation is not limited to just the spirals. This may be an indication of jet emission in these systems. However, the correlation could also be a result of the radio-IR correlation seen in SFGs due to SFA~\citep{2021A&A...647A.123D}. To test this hypothesis further would require the total radio luminosity for the entire galaxy rather than just the nuclear region, which is beyond the scope of this work. 


\begin{figure}
\begin{center}
Radio Correlation    \\
\includegraphics[width=0.44\textwidth]{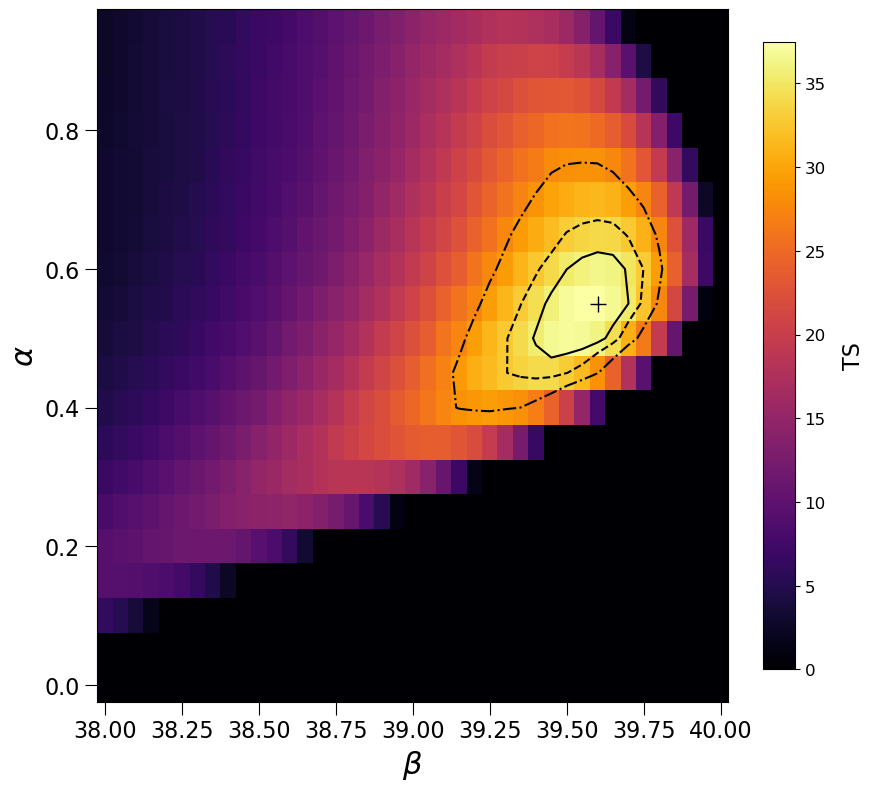} 

\caption{Stacked profile for the $\mathrm{L_{\gamma}- L_{15GHz}}$ correlation using a subset of 129 sources with available radio data. The color scale and contours are the same as described in Figure~\ref{fig:flux_index_stack}.}
\label{fig:AB_Lgam_L15GHz_all_spirals}
\end{center}
\end{figure}
From a qualitative point of view, 4/5 of the significantly detected sources are also classified as FR~I radio galaxies. Conversely, the subthreshold sample does not contain any radio galaxies, although compact jets have been identified in a number of the subthreshold sources. This is illustrated in Figure~\ref{TS_phind_morph}, which shows the photon index versus TS for the subthreshold sample (for sources with TS$\geq$4). The sources with jetted morphology have been identified from radio observations, as described in~\citet{2021MNRAS.500.4749B}. At least six galaxies in our subthreshold sample contain compact jets, although these sources show no obvious correlation in terms of TS or photon index, i.e., they are not uniformly the most significant of the subthreshold sources nor do they share a common $\gamma$-ray index.

\begin{figure}[t]
\begin{center}
\vspace{0.5cm}
\includegraphics[width=0.45\textwidth]{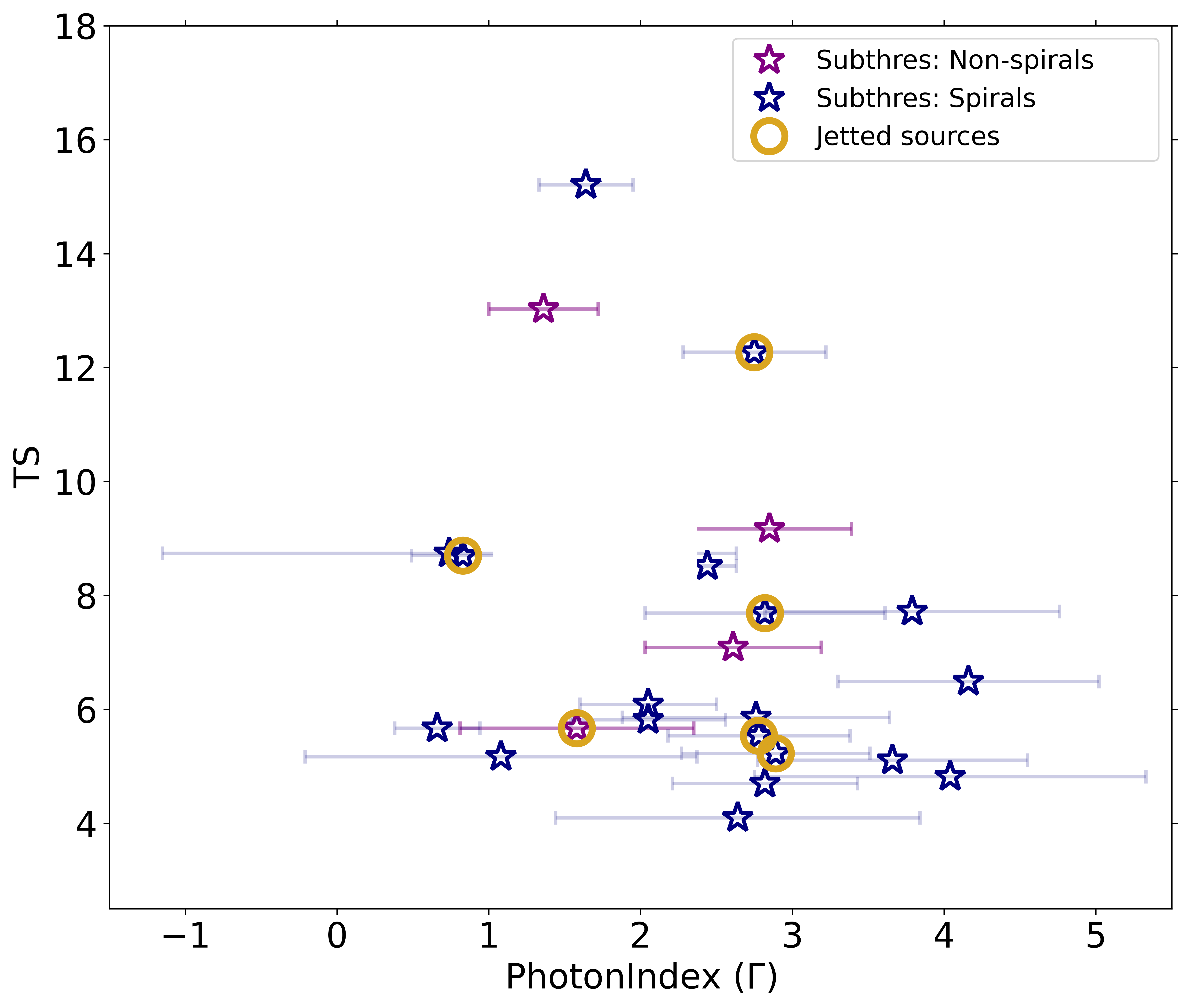}
 \caption{TS versus Photon Index ($\Gamma$), showing the subthreshold sources with their spiral and non-spiral classifications. The sources showing a jetted morphology in the radio band from~\citet{2021MNRAS.500.4749B} are denoted by yellow open circles.}
\label{TS_phind_morph}
\end{center}
\end{figure}



 

Interestingly, we note that our sample includes NGC 4278, which was recently associated with a TeV source by the Large High Altitude Air Shower Observatory (LHAASO) collaboration~\citep{2024ApJ...971L..45C}. Using the same time frame as the LHAASO campaign, the source was also recently detected by the LAT, where it was argued that the signal likely has a jet origin~\citep{2024ApJ...977L..16B,Dominguez:2025nxa}. In fact, this source has the second highest TS in the subthreshold sample (TS=13), with a best-fit spectral index of $1.4 \pm 0.4$, compatible with the hard spectrum found in~\citet{2024ApJ...977L..16B}. This result adds to the ambiguity of the radio correlation study, and also highlights the fact that the LLAGN sample is highly non-homogeneous. 

\subsection{Emission from Star Formation Activity and Compact Jets}
To directly test the hypothesis that the $\gamma$-ray signal has contributions from both SFA and compact jets, we implement a three-component stacking analysis. This method tests the luminosity correlation of the form:
\begin{equation}
\begin{split}
    \mathrm{log}L_\gamma = \alpha\mathrm{log}\bigg(\frac{L_{15\mathrm{GHz}}}{10^{36.9} \mathrm{\ erg \ s^{-1}}}\bigg) 
    \\ + \beta\mathrm{log}\bigg(\frac{L_{8-1000 \mu m}}{10^{43.3} \mathrm{\ erg \ s^{-1}}}\bigg) + \delta
\end{split}
\end{equation}
The parameters $\alpha$ and $\beta$ are scanned from 0 to 3 in steps of 0.1, and $\delta$ is scanned from 38 to 41 with a step size of 0.1. Since this test requires both $L_{15\mathrm{GHz}}$ and $L_{8-1000 \mu \mathrm{m}}$ data, we are left with 120 sources in the subthreshold sample. The resulting best-fit parameters from the fit are $\alpha$ = $0.4^{+0.1}_{-0.2}$, $\beta$ = $0.3^{+0.2}_{-0.2}$, and $\delta$ = $39.1^{+0.2}_{-0.2}$, with a maximum TS of 49.6 (6.5\,$\sigma$). Compared to the $L_{\gamma}- L_{IR}$ correlation, there is a slight improvement in TS for the three-component stacking ($\Delta$ TS $= 4.7$, $\sigma=1.9$ for 1 additional degree of freedom). Thus there is some indication that the signal may have contributions from both SFA and compact jets, but at this time we are unable to make any robust conclusions due to the low significance. Overall, given the ambiguity of the radio correlation study and the general non-homogeneity of the LLAGN sample, at this time we cannot uniquely determine the origin of the observed $\gamma$-ray signal from the subthreshold sample.

\begin{figure*}
\begin{center}
\includegraphics[width=0.47\textwidth]{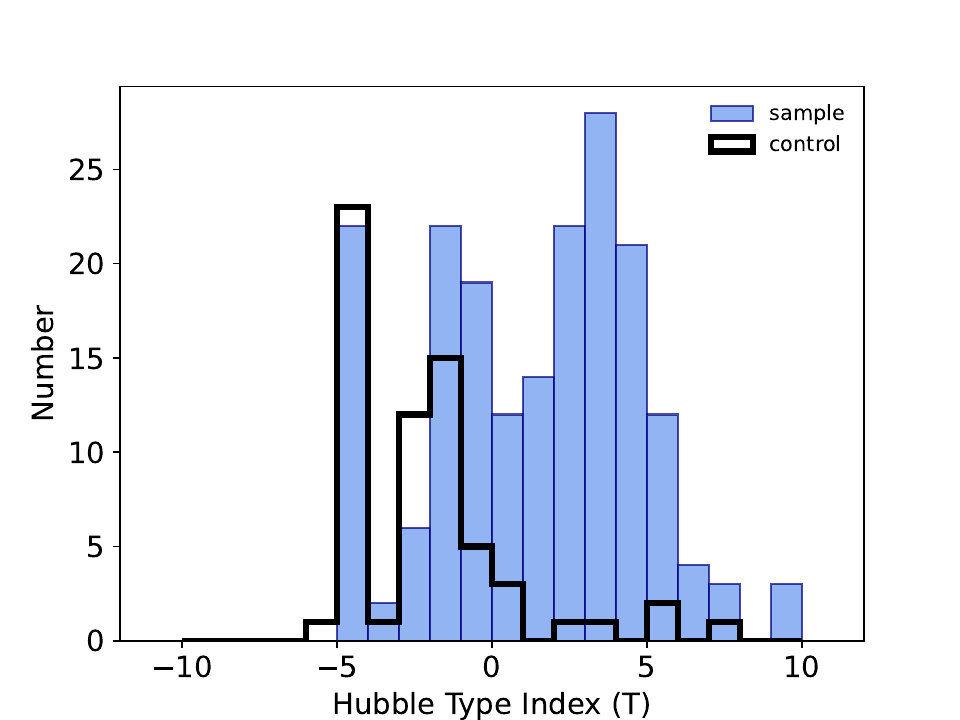} 
\includegraphics[width=0.47\textwidth]{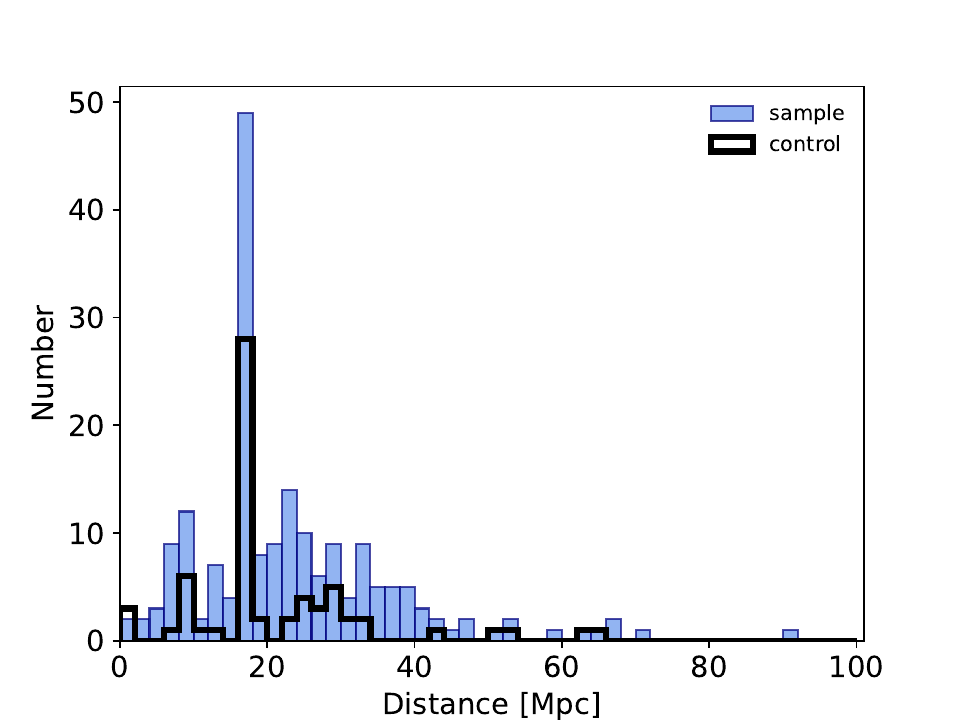} 
\includegraphics[width=0.47\textwidth]{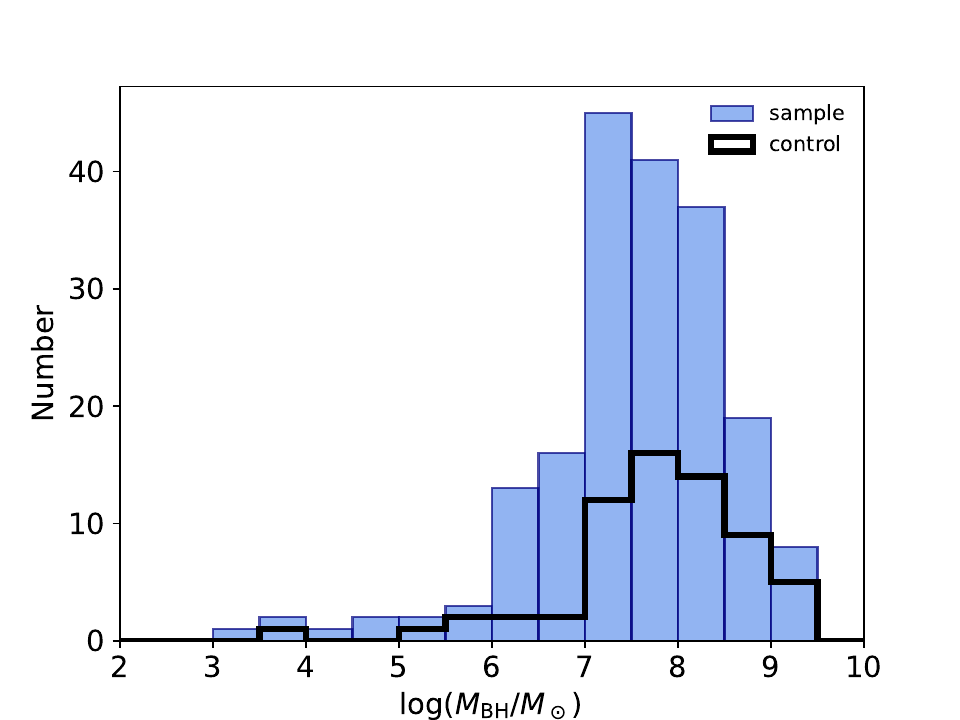} 
\includegraphics[width=0.47\textwidth]{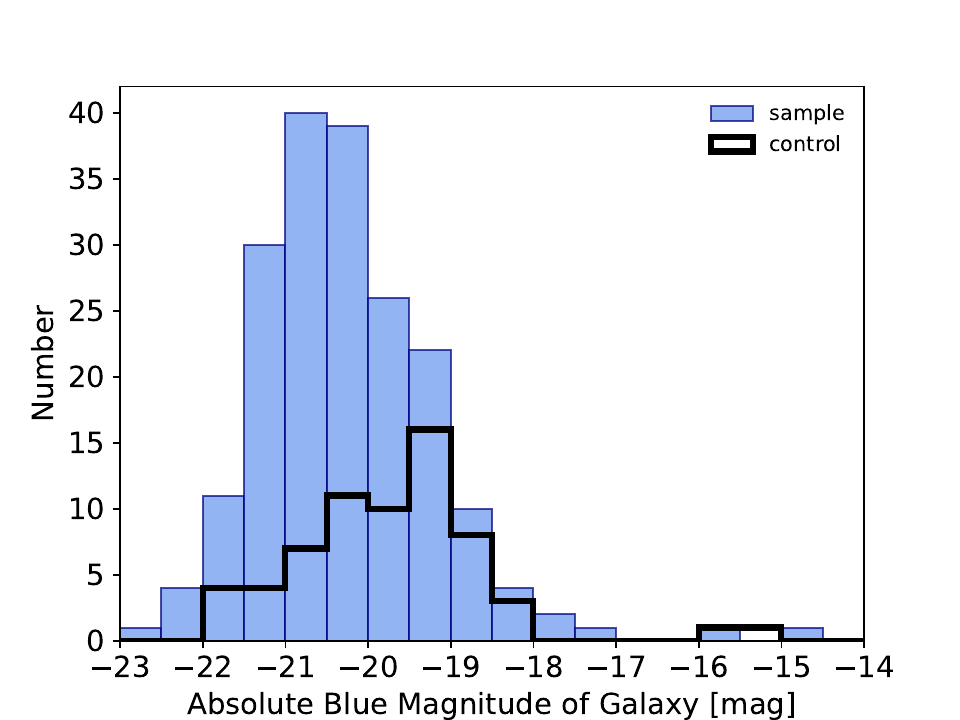} 
\includegraphics[width=0.47\textwidth]{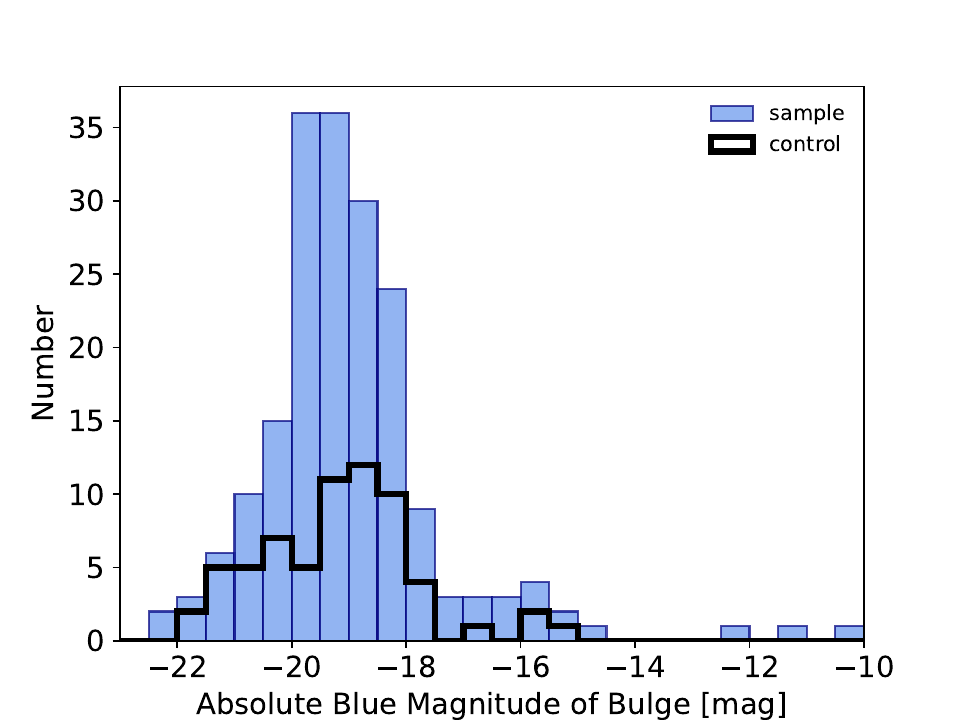} 
\includegraphics[width=0.47\textwidth]{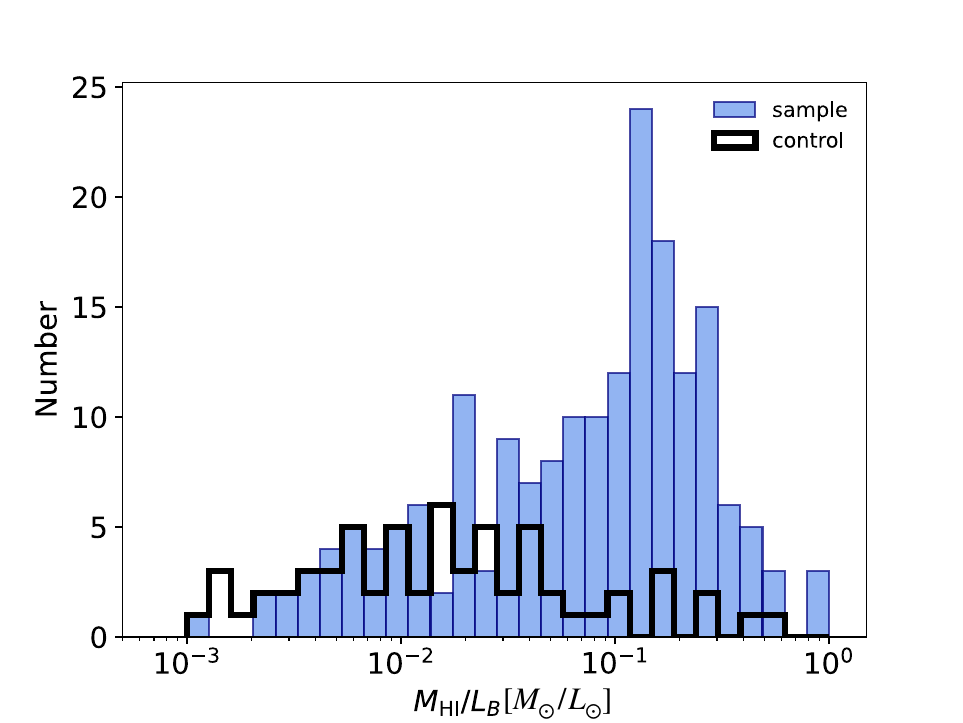} 
\caption{Comparison of different source properties for the control sample and subthreshold LLAGN sample. \textbf{Upper left:} Hubble Type index; \textbf{Upper right:} distance; \textbf{Middle left:} BH mass, normalized to the mass of the Sun; \textbf{Middle right:} Absolute blue magnitude of the entire galaxy (assuming the given distance given in the upper right), corrected for Galactic and internal extinction; \textbf{Bottom left:} Absolute blue magnitude of bulge, estimated from the total luminosity; \textbf{Bottom right:} HI mass normalized to the extinction-corrected blue-band luminosity.}
\label{fig:control}
\end{center}
\end{figure*}

\subsection{Control Sample}
\label{sec:controls}
Of the 486 total galaxies in the Palomar sample, 65 are free of any visible emission lines. We use these galaxies as a control sample. Figure~\ref{fig:control} compares the control sample and the subthreshold LLAGN sample, for different fundamental properties, which include Hubble Type index, distance, BH mass, absolute blue magnitude of the entire galaxy (includes galaxy and bulge), absolute blue magnitude of bulge, and atomic gas. This data is provided as part of the original Palomar survey and is described in~\citet{1997ApJS..112..315H}. We find that the samples are well matched in distance and BH mass, although they are not well matched in Hubble Type index, absolute blue magnitude, and H\,{\sc i} mass. 

The Hubble Type index quantifies the morphology of the galaxy. In general, indices $> 0$ are spirals, whereas values $\leq0$ are non-spirals. We can see that the control sample contains mostly non-spirals. This is in contrast to the subthreshold sample which contains both spirals and non-spirals. In fact, as already discussed, essentially all of the $\gamma$-ray signal is from the subset of spiral galaxies. Therefore, this control sample provides a limited point of comparison, since its morphology distribution differs substantially from that of the subthreshold sample. 

In Figure~\ref{fig:control}, the absolute blue magnitude of the entire galaxy (assuming the given distance) has been corrected for Galactic and internal extinction, and it serves primarily as a tracer of recent star formation and the presence of young, massive stars. Compared to the control sample, the LLAGN sample is shifted towards lower magnitudes (meaning higher B-band luminosities). The figure also shows the absolute blue magnitude of the bulge, which is estimated from the total luminosity (disk + bulge). The mass of the bulge most likely has the strongest impact on the nuclear activity. Again, we note that the bulges of the sample galaxies have fainter magnitudes on average than the control sample, although not as apparent as for the entire galaxy. 

The H\,{\sc i} mass is normalized to the extinction-corrected blue-band luminosity. This gives an indication of the availability of interstellar matter, which should affect the fueling rate of the nuclear regions, as well as the fueling rate of SFA. The LLAGN sample has significantly more galaxies with higher H\,{\sc i} mass, indicating that they have higher fueling rates.

\begin{figure}
\begin{center}
Control Sample
\includegraphics[width=0.45\textwidth]{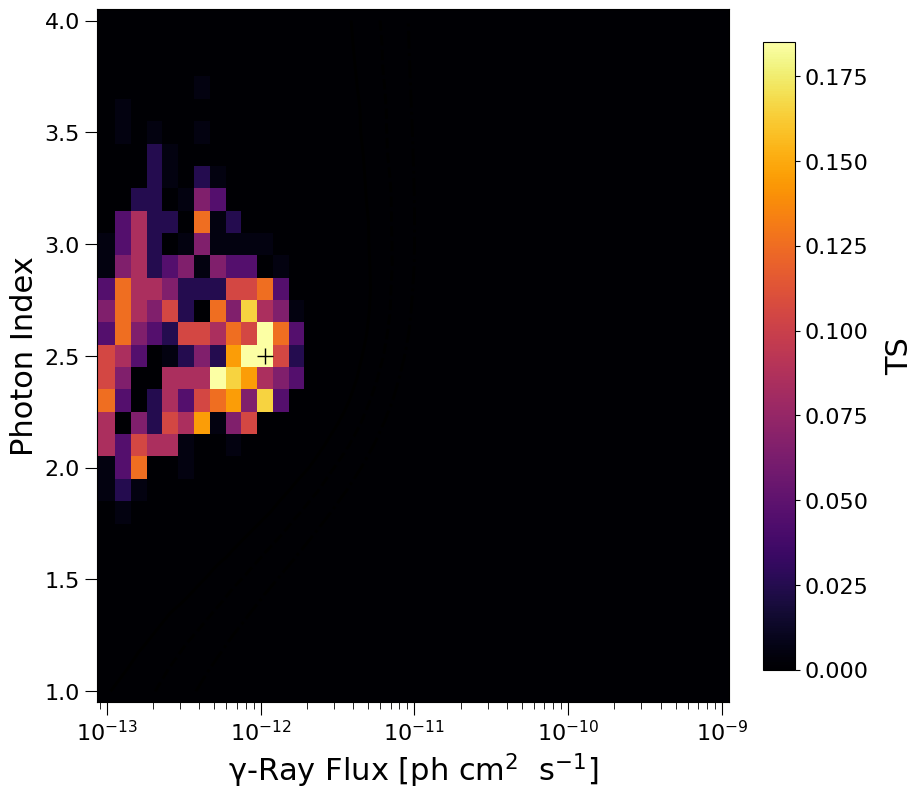} 
\caption{Stacked profile for the control sample. The color scale shows the TS, and the plus sign indicates the location of the maximum value, with a TS = 0.2.}
\label{fig:control-stack}
\end{center}
\end{figure}
We repeat the stacking analysis for the control sample, in the same way as it is performed for the nominal LLAGN sample. Results for the stacked profile are shown in Figure~\ref{fig:control-stack}. We find no signal, with a maximum TS of 0.2. This null result further supports the hypothesis that the observed $\gamma$-ray signal is physically associated with the sample of LLAGN.

\section{Model and Interpretation}
\label{sec:model}

\begin{figure*} 
\begin{center}
\includegraphics[width=0.4\textwidth]{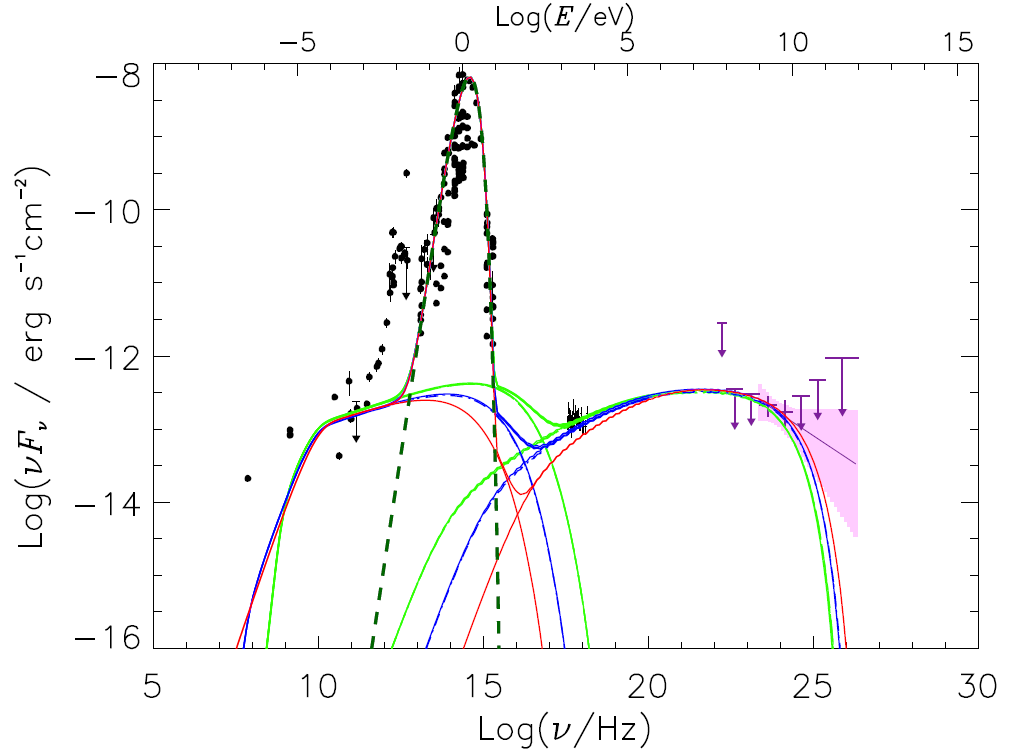} 
\includegraphics[width=0.4\textwidth]{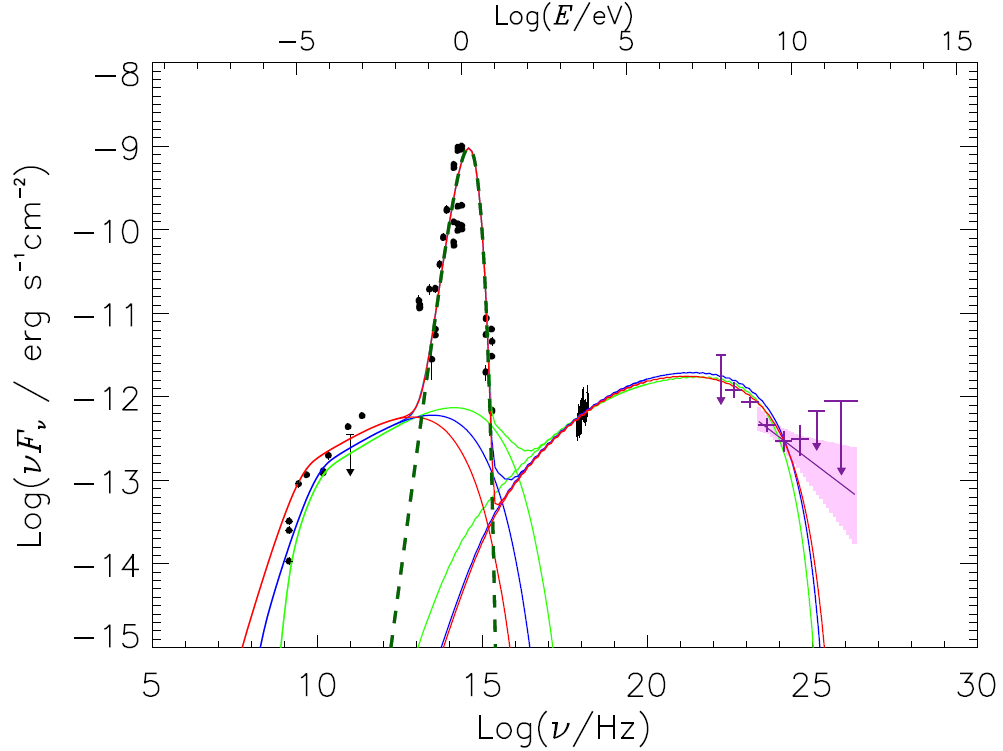} 
\includegraphics[width=0.4\textwidth]{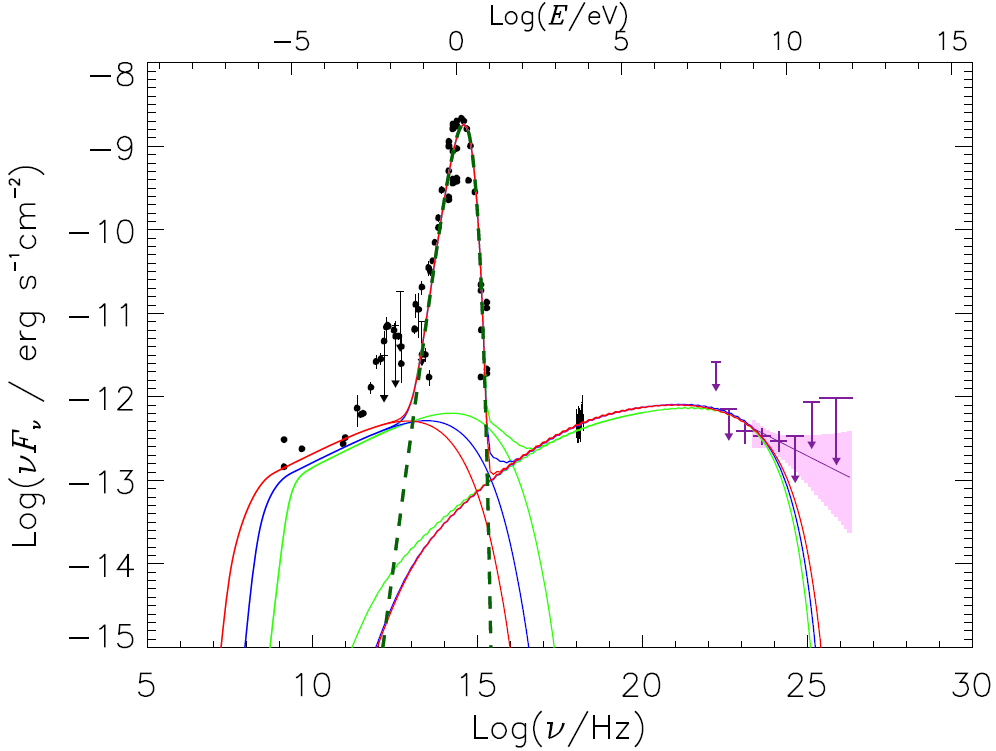} 
\caption{Broadband SEDs for M84 (upper-left), NGC 315 (upper-right), and NGC~4261 (lower), represented by one-zone steady-state SSC models
with emitting exponential cutoff power law particle spectra, with indices and Lorentz factors of $p = 2.7$, $\Gamma = 1.02$; $p = 2.5$, $\Gamma = 1.0001$; and $p = 2.6$, $\Gamma = 1.0001$; respectively. The sources are constrained within 
the jet power limit of $L_{\rm jet} \lesssim 10^{43.3}$ erg s$^{-1}$ (green
curves: $R = 10^2r_g$, blue curves: $R = 10^3r_g$, red
curves: $R = 10^4r_g$). Data from the LAT is shown in violet, and archival data from other wavebands is shown in black. Errors on the data are at the 1\,$\sigma$ level, and upper limits are at the $95\%$ confidence level. The dashed line represents the host galaxy’s SED.
}
\label{fig:model_SEDS}
\end{center}
\end{figure*}

One of the goals of this work is to study the origin of the high-energy processes in LLAGN. For the subthreshold sample, we find that the $\gamma$-ray emission is consistent with being dominated by SFA, which prevents us from uniquely probing the disks and jets in these systems. We therefore concentrate our modeling on a subset of the individually detected galaxies -- NGC~315: redshift $z = 0.0165$, $M_{BH} = 2.08\times 10^9 M_\odot$ \citep{2021ApJ...908...19B}; NGC~4261 (C~270): $z = 0.007261$, $M_{BH} = 1.62\times 10^9 M_\odot$, e.g.,~\citet{2023MNRAS.522.6170R}; and NGC~4374 (M84): $z=0.003392$, $M_{BH} = 8.5\times 10^8 M_\odot$, e.g.,~\citet{2010ApJ...721..762W}. All of these are well known radio galaxies of type FR~I, their nuclear regions classified as LINERs, with early-type host galaxies. The two individually detected galaxies not included in our modeling are very well known sources---NGC 4486 (commonly known as M87) and NGC 1275---and have already been well explained in terms of synchrotron and SSC jet emission~\citep{2009ApJ...707...55A,2009ApJ...699...31A}. Broadband SEDs for the subset of galaxies are shown in Figure~\ref{fig:model_SEDS}, where we include \textit{Fermi}-LAT data from this work, archival data from other wavebands, and our SED modeling, as discussed below. 

The archival data spanning radio to optical bands is taken from the NASA Extragalactic Database (NED)\footnote{\url{https://ned.ipac.caltech.edu/}}, accessed on 2024 May 14~\citep{ned}. In order to ensure that the SED modeling focuses on the jet emission, we perform several cuts on the data. First, we discarded data coming from observations not directly pointing at the sources (i.e., our source of interest is in the field of view but the pointing is centered at a different source). Furthermore, we do not include data for which no flux uncertainty is reported, as well as data for which the associated comments state that no corrections have been made (e.g.,~either for known sources in the field of view, or for galactic extinction). Additionally, we discard data provided with a poor quality flag or explicitly stating that the observation was of a component of the AGN other than the jet (e.g., the nuclear environment or a thermal component).

The X-ray spectra of the three sources were extracted from their longest archival Chandra observations, where there are 46~ks (ObsID: 5908), 55~ks (ObsID: 4156), and 101~ks (ObsID: 9569) for M84, NGC~315, and NGC~4261, respectively. The spectra were extracted with {\tt CIAO} version 4.15, using a circular region with a radius of 1\arcsec. The background spectra were extracted from circular regions with a radius of 15\arcsec \ that are 2.5\arcmin, 1.5\arcmin, and 1.5\arcmin \ away from the source regions of M84, NGC~315, and NGC~4261, respectively. The analysis was performed with {\tt XSPEC} version 12.13.1~\citep{1996ASPC..101...17A} using $\chi^2$ statistics in the $0.5-7$ keV energy range. The spectra were binned with a minimum of 20 counts per bin using the HEAsoft tool {\tt grppha}. The X-ray spectrum of M84 was modeled using an absorbed power law with a best-fit photon index of $\Gamma = 2.0 \pm 0.1$ and a best-fit column density of $\mathrm{N_H}\sim 1.4 \times 10^{21} \ \mathrm{cm^{-2}}$. The X-ray spectrum of NGC~315 was modeled with two components, including an APEC component (with a best-fit temperature of $\sim 0.54$ keV) to model the soft excess and an absorbed power law to model the absorbed intrinsic emission, with a best-fit photon index of $\Gamma = 1.5 \pm 0.1$ and a best-fit column density of $\mathrm{N_H} \sim 1 \times 10^{22} \ \mathrm{cm^{-2}}$. The X-ray spectrum of NGC~4261 was modeled with three components following~\citet{2010MNRAS.408..701W}, using an APEC component (with a best-fit temperature of $\sim 0.66$ keV) to model the soft excess; an absorbed power law to model the absorbed intrinsic emission, with a best-fit photon index of $\Gamma = 1.3 \pm 0.3$ and a best-fit column density of $\mathrm{N_H}\sim 9 \times 10 ^{22} \ \mathrm{cm^{-2}}$; and an unabsorbed power law to model the leaked intrinsic emission from the center of the AGN (assuming the same photon index as applied in the absorbed power law component). We used the unabsorbed intrinsic power law component spectra when performing the SED fitting of the three sources.

As for many early-type radio-loud galaxies, these sources show strong dust emission, in addition to a central gas emission disk. The central dust morphologies range from disks (NGC~315; NGC~4261's small nuclear dust disk of $\sim$ 240\,pc in diameter has been suggested to possess rather a spiral structure) to lanes (M84 has 2 warped dust lanes of $\sim 1$ kpc in size)~\citep{1999AJ....118.2592V}. Furthermore, a jet-like feature ($\sim 90$ pc in length) of dust protruding from the dust disk and pointing away from the nucleus, and a somewhat weaker similar feature in the counter-jet direction, have been detected in NGC~4261's central region. \citet{1999AJ....118.2592V} report several additional patches of dust on either side of NGC~315's dust disk. Unsurprisingly, the SEDs from the central regions of all three sources show clear signatures of excess IR emission (with respect to our simplified input model) that we attribute to dust features. For the present work, aiming towards an understanding of the sources’ non-thermal emission, we leave this IR excess emission unmodeled. The host galaxies’ optical emissions are represented (dashed lines in Figure~\ref{fig:model_SEDS}) by gray-body thermal spectra with temperature $T\sim 9000$ K, similar to the work of~\citet{2024ApJ...971...84K}.

The kinematics of the two-sided inner jets of all three sources have been studied extensively in the radio domain using VLBI/VLBA, revealing valuable constraints for our jet emission modeling.  \citet{2022ApJ...941..140W} derive moderately relativistic intrinsic speeds of 3 components in the inner jet of M84 based on several epochs of observations in 2019--2021, adding to historical data of 2014. The combined analysis revealed an accelerating inner jet (intrinsic speeds of $0.12c$, $0.27c$, $0.32c$) at average de-projected distances from the core of $1960~r_g$, $4200~r_g$ and $7400~r_g$, respectively, with a gravitational radius, $r_g = 1.25\times 10^{14}$ cm, for its (inner) jet viewing angle $\theta$ of $58^{+17}_{-18}\deg$. Likewise, proper motion measurements from \citet{2023ApJ...957...32Y} of several emission components in the approaching and receding jets of NGC~4261 revealed accelerating, mildly relativistic (parabolic-shaped until around $8\times 10^3~r_g$, with $r_g =  2.38\times 10^{14}$ cm, where the collimated flow transitions to a conical shaped one) bulk flows with intrinsic speeds between $0.3c - 0.55c$ at a jet viewing angle between $54\deg$ and $84\deg$. Similarly, the modeling of NGC~315's inner jet VLBI data of \citet{2021A&A...647A..67B}(see also \citet{2021ApJ...909...76P}) by \citet{2022A&A...664A.166R} revealed an accelerating parabolic-shaped  jet on sub-pc scales with a maximum average viewing angle of $44\pm4\deg$, and the highest bulk velocity of $\sim 0.95c\pm 0.03c$ reached at around 1~pc (corresponding to $\sim 10^4 r_g$ with $r_g = 3.07\times 10^{14}$ cm) from the core.

In line with our goal to examine whether a simple steady-state one-zone leptonic jet emission model is suitable to broadly represent the broadband radio-to-$\gamma$-ray SED, we use a simplified geometry and bulk equation of motion for its emission region: a spherical, homogeneously magnetized (with magnetic field strength $B$) region of co-moving size $R_{\rm jet}$ containing relativistic pairs (referred to as electrons in the following) of co-moving energy density $u_e$, and the same number of cold protons of co-moving energy density $u_p$, which moves with constant bulk speed along the jet axis. We note that mounting evidence supports the need for protons and/or ions in jets of AGN~\citep[see, e.g.,][]{2006ApJ...648..200D,2008MNRAS.385..283C}, although their properties are still debated. To account for this need, we use cold protons in our emission modeling of a charge-neutralized jet. For the jet emission modeling (using the same radiation model as in~\citet{2024ApJ...971...84K}) of the three sources, we fix their viewing angles at the values implied by the interferometric radio measurements ($58\deg$, $63\deg$ and $44\deg$ for M84, NGC~4261 and NGC~315, respectively), while scanning the bulk motion Lorentz factors $\Gamma$ from $\sim 1$ to 1.5, corresponding to the overall moderately relativistic intrinsic speeds, $\beta c = c\sqrt{1-\Gamma^{-2}}$, implied by the aforementioned sub-pc data. We require the corresponding jet power to lie below $2\times 10^{43}$ erg/s, computed for each parameter set through
\begin{equation}
    L_{\mathrm jet}=4\pi c R_{\mathrm jet}^2 \beta^2\Gamma^2 (u_B + u_{\mathrm{particles}}),
\end{equation}
with $u_B$ the magnetic field energy density. This estimate for the maximum (inner) jet power is motivated by the measurement of X-ray cavities as reported by \citet{2006MNRAS.372...21A} for M84, $(1.53\pm0.46)\times 10^{43}$ erg/s, and the morphological similarity of all three jet sources.

Despite the prominent IR thermal emission component observed in all three sources, its role as a target photon field for inverse Compton scattering off relativistic jet electrons can be neglected, mainly due to the much larger spatial size (of order $\sim 100$ pc) of this dust component as compared to the jet emission regions: the ratio of this IR component’s target photon energy density in the jet co-moving frame, $u_{\mathrm IR}$, to the co-moving jet photon energy density, $u_{\mathrm jet}$, can be estimated from
\begin{align}
\frac{u_{\mathrm{IR}}}{u_{\mathrm{jet}}} &\approx 10^{-8} \left(\frac{F_{\mathrm{obs,IR}}}{F_{\mathrm{obs,jet}}}\right) \nonumber \\
&\quad \times \left(\frac{R_{\mathrm{jet},16}}{R^*_{\mathrm{IR},20}}\right)^2 D^4 \Gamma^2.
\end{align}
Here, $R_{\mathrm jet} = 10^{16} R_{\mathrm jet,16}$ cm, $R^*_{\mathrm IR} = 10^{20} R^*_{\mathrm IR,20}$ cm is the size of the thermal IR source in the AGN frame, $F_{\mathrm obs,jet}=(R_{\mathrm jet}/d_l)^2 c u_{\mathrm jet} D^4$ and $F_{\mathrm obs,IR}=(R^*_{\mathrm IR}/d_l)^2 c u_{\mathrm IR} \Gamma^{-2}$ are the observed $\nu F_\nu$ jet synchrotron and thermal IR flux, respectively. With the Doppler factor $D\approx 1\approx \Gamma$  for the three considered sources, one then finds $u_{\mathrm IR}\ll u_{\mathrm jet}$. The photon energy densities of the (assumed kpc-sized) host galaxies of the three LLAGN are estimated to be $\sim 10^{-10}$ erg/cm$^3$, or $\sim10^{-10} \Gamma^2$ erg/cm$^3$ in the jet co-moving frame. This compares to the co-moving jet photon energy density $u_{\mathrm{jet}}$ as $u_{\mathrm{gal}} \approx 10^{-11} u_{\mathrm{jet}} F_{\mathrm{obs,gal}}/F_{\mathrm{obs,jet}}\times (R_{\mathrm{jet},16}/R^*_{\mathrm{gal,kpc}})^2 D^4 \Gamma^2 \ll u_{\mathrm{jet}}$. We note that while the dust and host galaxy radiation fields can thus be neglected as targets for particle-photon interactions for the rather compact nuclear emission regions considered here, they may become relevant to the production of any $\gamma$-ray flux in large-scale ($\gtrsim$ tens kpc) extended jets~\citep[e.g.,][]{2003ApJ...597..186S}. As is typical for FR~I sources, other putative photon fields external to the jet such as the accretion disk and broad-line region can be neglected as targets for jet-electron interactions, leaving the synchrotron self-Compton (SSC) process as the dominant $\gamma$-ray production channel for a leptonic emission model.

As in~\citet{2024ApJ...971...84K}, a relativistic electron distribution with power law index $p$ between electron Lorentz factor $\gamma_{\rm min}$ and exponential cutoff $\propto \exp(-\gamma/\gamma_{\mathrm{max}})$ underlies the synchrotron and SSC emission components of the computed SEDs. The X-ray spectral shapes measured from the three sources suggest a SSC interpretation, which we use for our modeling. We note that the joint X-ray and LAT spectral measurements provide severe constraints on the flux level of the SSC component within the framework considered here. By scanning through a range of region sizes, $R_{\rm jet}$, and adjusting magnetic field strengths and jet particle densities to provide similar levels of synchrotron power, we find examples of broadband SEDs that represent the data well, as shown in Figure~\ref{fig:model_SEDS}.

Only very slow jets ($\beta < 0.2-0.3$) for emission region sizes $R_{\rm jet}\sim 10^{2-4}r_g$ are found to meet all observational (including jet power) constraints of the three considered FR~I sources. With $\gamma_{\rm max} \sim (7-20) \times 10^4$ (M84),  $\gamma_{\rm max} \sim (2-6) \times 10^4$ (NGC~315),  $\gamma_{\rm max} \sim (3-8) \times 10^4$ (NGC~4261) and field strengths of $B\sim 0.002-0.3$ G (M84), $B\sim 0.003-0.4$ G (NGC~315, NGC~4261), the synchrotron components only reach the optical to (hard) UV band, and the SSC components cut off at GeV-energies, with comparable power in both components. Within the framework of leptonic-only radiation models, these sources are therefore predicted to be unfavorable for detection at TeV energies. The overall jet composition is found to be strongly particle dominated ($u_B/(u_e+u_p) \sim (0.1-3)\times 10^{-3}$).

To summarize, the $\gamma$-ray radiation from the three considered strongly misaligned FR~I LLAGN can be understood as inverse Compton scattered jet synchrotron photons, if the (inner) jet emission region is weakly magnetized with its total energy density being strongly particle dominated, and only slowly moving. 

\section{Summary and Conclusion}
\label{sec:discussion_and_conclusion}
We have presented a comprehensive analysis of the $\gamma$-ray emission from LLAGN in the Palomar sample using 14.4 years of \textit{Fermi}-LAT data. Our stacking analysis of 186 subthreshold sources yields a significant detection (5.2\,$\sigma$), with the signal largely driven by spiral galaxies. We find strong evidence that the observed emission correlates with infrared luminosity, consistent with SFA as the dominant source. However, a correlation with the core radio luminosity is also present, which may indicate a contribution from compact jets in at least some of these systems. On the other hand, this trend could also arise from the radio–IR correlation in SFGs due to SFA. Thus, the current results do not allow us to uniquely determine the origin of the subthreshold signal. 

We report a new significant LAT detection of NGC 4374, bringing the number of firmly detected LLAGN to five. For three of these sources--NGC 315, NGC 4261, and NGC 4374--we perform detailed broadband spectral modeling. We show that their $\gamma$-ray emission can be explained by synchrotron self-Compton emission from weakly magnetized, slowly moving, and particle-dominated jet regions.

These findings suggest a complex interplay between star-formation and jet-related processes in LLAGN and demonstrate the value of deep LAT observations combined with stacking techniques for studying faint AGN populations. Our study also highlights the fact that the Palomar sample of LLAGN is highly non-homogeneous, complicating any attempt to generalize the $\gamma$-ray properties. Our publicly released stacking library will support future studies of subthreshold sources across a variety of astrophysical contexts.

 

\section*{Acknowledgements}
C.~M.~K.~and M.~A.~acknowledge funding under NASA Contract No.~80NSSC22K1580 (Fermi Guest Investigator Program Cycle 15 No.~151048). C.~M.~K.'s research was supported by an appointment to the NASA Postdoctoral Program at NASA Goddard Space Flight Center, administered by Oak Ridge Associated Universities under contract with NASA.

This work was supported by a grant from the Simons Foundation International [SFI-MPS-SSRFA-00012817, C.~M.~K].

Clemson University is acknowledged for its generous allotment of compute time on the Palmetto Cluster.

The \textit{Fermi}-LAT Collaboration acknowledges generous ongoing support
from a number of agencies and institutes that have supported both the
development and the operation of the LAT as well as scientific data analysis.
These include the National Aeronautics and Space Administration and the
Department of Energy in the United States, the Commissariat \`a l’Energie Atomique
and the Centre National de la Recherche Scientifique / Institut National de Physique
Nucl\`eaire et de Physique des Particules in France, the Agenzia Spaziale Italiana
and the Istituto Nazionale di Fisica Nucleare in Italy, the Ministry of Education,
Culture, Sports, Science and Technology (MEXT), High Energy Accelerator Research
Organization (KEK) and Japan Aerospace Exploration Agency (JAXA) in Japan, and
the K.~A.~Wallenberg Foundation, the Swedish Research Council, and the
Swedish National Space Board in Sweden.

Additional support for science analysis during the operations phase is gratefully
acknowledged by the Istituto Nazionale di Astrofisica in Italy and the Centre
National d’\`Etudes Spatiales in France. This work performed in part under the DOE
Contract DE-AC02-76SF00515.

\bibliography{LLAGN_citations}{}
\bibliographystyle{aasjournal}

\end{document}